\documentclass[aps,jcp,twocolumn,superscriptaddress,10pt]{revtex4-2}

\usepackage[T1]{fontenc}
\usepackage[utf8]{inputenc}
\usepackage[version=3]{mhchem}
\usepackage{amssymb}
\usepackage{xcolor}
\usepackage{braket}
\usepackage{siunitx}
\usepackage{graphicx}
\usepackage[normalem]{ulem}

\begin{document}

\title{First-order Degenerate Symmetry-Adapted Perturbation Theory}

\author{Dominik Cieśliński}
\email{d.cieslinski@uw.edu.pl}
\affiliation{University of Warsaw, Faculty of Chemistry, Pasteura 1, 02-093 Warsaw, Poland}

\author{Tijs Karman}
\affiliation{Institute for Molecules and Materials, Heijendaalseweg 135, 6525 AJ Nijmegen, Radboud University, The Netherlands}

\author{Anthony Scemama}
\affiliation{Laboratoire de Chimie et Physique Quantiques - UMR5626, CNRS/Université Paul Sabatier, Bat. 3R1b4, 118 route de Narbonne, 31062 Toulouse Cedex 09, France}

\author{Michał Hapka}
\email{michal.hapka@uw.edu.pl}
\affiliation{University of Warsaw, Faculty of Chemistry, Pasteura 1, 02-093 Warsaw, Poland}

\author{Piotr S.\ {\.Z}uchowski}
\affiliation{Faculty of Physics, Astronomy and Informatics, Grudziadzka 5, 87-100 Torun, Nicolaus Copernicus University, Poland}

\begin{abstract}
Degenerate electronic states govern the photophysics of excimers and the molecular dynamics near conical intersections. In regions where potential energy surfaces are degenerate or nearly degenerate, intermolecular interactions mix the corresponding electronic configurations. We introduce a first-order degenerate formulation of symmetry-adapted perturbation theory (dSAPT) with weak symmetry forcing that quantifies the mixing in terms of electrostatic and exchange (Pauli repulsion) contributions. We compare two alternative ways of resolving the degeneracy, in which antisymmetry is enforced either \textit{after} or \textit{before} diagonalization of the first-order eigenproblem. Using CASSCF monomer wave functions, we apply dSAPT to two model spatially degenerate systems, \ce{H2}-F(${}^2$P) and \ce{H2}-NO(${}^2\Pi$), and show that exchange effects contribute significantly to configuration mixing already at intermediate intermolecular separation. For the \ce{H2}-NO(${}^2\Pi$) complex, we further demonstrate that second-order dispersion energy is essential for qualitatively reproducing the angular dependence of the mixing (diabatic) angle. Finally, using water and benzene excimers with monomers described either with CASSCF or CIS wave functions, we show that Pauli repulsion controls delocalization of the excitation at short intermolecular separations.
\end{abstract}
 

\maketitle

\section{Introduction}

Since its early implementations, symmetry-adapted perturbation theory (SAPT) has matured into a standard framework for studying molecular interactions \cite{jeziorski1994per,patkowski2020rec}. In addition to providing accuracy which can be competitive with gold standard coupled cluster methods \cite{Parker:14,Masumian:24}, SAPT has repeatedly contributed to the quantitative interpretation of intuitive concepts such as hydrogen or halogen bonding \cite{hapka2026sym}. This interpretability stems from the fact that individual orders of the perturbation series naturally represent physically meaningful energy components expressible in terms of monomer properties (density matrices and response functions).

Recently, the development of SAPT has progressed toward states with multireference character. Patkowski and co-workers \cite{patkowski2018fir,waldrop2019spi} introduced a generalization of SAPT for cases with spin degeneracy, i.e., two open-shell subsystems that couple to arbitrary spin multiplicities. This approach enables direct calculations of the spin-exchange energy, with  good overall performance demonstrated for systems such as the oxygen dimer and phenalenyl dimer. In parallel, multiconfigurational SAPT \cite{Hapka:19a,hapka2019sec,hapka2021sym} has allowed to describe noncovalent complexes where at least one monomer requires a multideterminant reference, for instance when static correlation effects cannot be neglected or when targeting excited states with localized excitations \cite{jangrouei2022dis,hapka2023eff,krzeminska2024ani,cieslinski2025fir}.

These multireference SAPT formulations share a common assumption: the zeroth-order states of the monomers are energetically well separated from other states. When the interaction strength becomes comparable to the spacing between monomer levels, this assumption collapses. The interaction mixes the states, and a nondegenerate perturbation expansion loses its meaning. This regime is by no means exotic, as it lies at the heart of photochemistry and nonadiabatic dynamics: it arises whenever potential energy surfaces approach one another near conical intersections. It also defines excimers, in which an excitation shared between two identical molecules makes the $\ket{A^*B}$ and $\ket{AB^*}$ configurations exactly degenerate. From the standpoint of degenerate perturbation theory, even when no symmetry operation enforces an equal-weight superposition of the two excitation-localized configurations, the states remain degenerate in the noninteracting limit, and their mixing must be resolved perturbatively.

Addressing noncovalent interactions involving degeneracies via supermolecular approach is challenging. Multireference methods demand a careful choice of the active space, and their computational cost grows factorially with the number of active orbitals. Accounting for dynamical correlation, which is essential for van der Waals forces, introduces further complications: multireference configuration interaction (MRCI) requires size-consistency corrections \cite{davidson1977siz}, whereas multireference perturbation theory and coupled-cluster methods frequently suffer from intruder-state problems \cite{paldus1993app,evangelista2018per}. A separate challenge is the need for a diabatic representation in dynamics simulations. Although various diabatization schemes have been proposed \cite{karman2018dia,shu2022dia}, constructing a globally consistent adiabatic-to-diabatic transformation remains a non-trivial task, especially when many electronic states are involved \cite{karman2016com}. Defining a proper counterpoise correction (CP) \cite{boys1970the} for the basis set superposition error (BSSE) is similarly complex. Among several generalizations of the CP procedure proposed for open-shell fragments, only the approaches of Alexander \cite{alexander1993adi} and Kłos et al. \cite{klos2001abi} recover the correct long-range behavior of the potential \cite{karman2018dia}. In excimers, standard CP corrections fail at short distances due to charge-transfer effects \cite{rocharinza2006ath,barcza2023ben}. More generally, low-lying Rydberg  states, common in small and medium-sized molecules, can strongly mix with valence states once diffuse basis functions are included \cite{barcza2023ben}. In larger excimers, the challenge shifts toward a reliable description of dispersion, which remains a major source of errors in TD-DFT employing ground-state dispersion corrections \cite{hancock2023non}.

Attempts to extend SAPT to electronically degenerate systems have a long history. In a seminal 1980 contribution, Chałasiński and Szalewicz \cite{chalbie1980deg} generalized several exchange perturbation theories \cite{murrell1967int,musher1967the,eisenschitz1930ube,avoird1967per,hirschfelder1967} to degenerate systems at arbitrary orders in the intermolecular interaction operator. Although the resulting power series expansions proved divergent for excited states of the model H$\cdots$H$^+$ ion, encouraging accuracy was obtained at low orders in the van der Waals minium region. Degenerate SAPT was revisited nearly two decades later by Korona et al. \cite{korona1999deg}, who proposed an alternative to the conventional power-series expansion based on wave operators \cite{bloch1958sur,lindgren1982per} and localization techniques \cite{williams1974loc,klein1974deg,chipman1977loc}. Using the interaction between the ground state helium atom and excited hydrogen atom as an example, they demonstrated satisfactory convergence near the ground-state van der Waals minimum, with the notable exception of the Rydberg C${}^2\Sigma^+$ state of HeH, where strong mixing with higher states of $\Sigma$ symmetry led to slow convergence. Interestingly, however, the earliest degenerate SAPT application was published in 1979 by van Hemert and van der Avoird \cite{vanHemert1979abi} who examined the lowest singlet state of the water excimer. Although their study was limited to the first-order interaction energy and the density-matrix expressions derived from the formalism of
Jeziorski et al.\ \cite{jeziorski1976fir} were specific to singly excited Slater determinants from the extended Hartree-Fock (electron-hole potential) method \cite{morokuma1972ext}, it remains the only many-electron degenerate SAPT application reported so far.

Despite these developments, SAPT has not evolved into a practical method for systems near crossings or close-lying regions of interacting potential energy surfaces, nor for interactions involving excitations delocalized over many subsystems. The so-called strong-symmetry forcing formulations explored in Refs.\citenum{chalbie1980deg} and \citenum{korona1999deg} have proven computationally insurmountable for monomers with more than several electrons, even in the nondegenerate case. Crucially, the variant of SAPT that can be generalized to many-electron systems, namely the symmetrized Rayleigh-Schr{\"o}dinger (SRS) \cite{jeziorski1978sym} theory, has not yet been extended to account for degeneracies.

In this work, we formulate first-order degenerate SAPT (dSAPT) within symmetrized Rayleigh–Schrödinger theory to treat (quasi)degenerate monomers in many-electron systems. We address a basic theoretical choice in degenerate SRS by defining and comparing two distinct antisymmetrization protocols: imposing antisymmetry either before or after resolving the electronic degeneracy. The general first-order equation, common to SRS and more involved exchange perturbation schemes \cite{chalbie1980deg}, was reported previously in Refs.~\citenum{vanHemert1979abi} and \citenum{chalbie1980deg}. However,  only the protocol in which antisymmetry is imposed before resolving the degeneracy was analyzed. Applying dSAPT to representative dimers that exhibit orbital (F$\cdots$H$_2$, NO$\cdots$H$_2$) and resonance-type (H$_2$O$\cdots$H$_2$O$^*$, benzene$\cdots$benzene$^*$) degeneracies, we determine how exchange coupling between potential energy surfaces modifies configuration mixing and exciton localization. We show that exchange can qualitatively alter the picture obtained from electrostatics alone already at intermediate intermonomer distances. For the open-shell dimers involving two diabatic states, we additionally assess the roles of second-order induction and dispersion using nondegenerate single-reference SAPT variants. 

The development of degenerate SAPT for excited states can be viewed in the broader context of methods designed to analyze exciton delocalization. Ge and Head-Gordon \cite{ge2018ene} extended the absolutely localized molecular orbital energy decomposition analysis (ALMO-EDA) \cite{khaliullin2007unr,horn2016pro} to treat resonance-type degeneracies at the CIS and TD-DFT levels. However, the resulting decomposition of the interaction energy into frozen, excitonic splitting, polarization, and charge transfer terms does not map directly \cite{narayan2026com} onto the electrostatics, exchange, induction, and dispersion partitioning of perturbation theory. In particular, the excitonic splitting term bundles together electrostatic, exchange, and higher-order couplings. Moreover, in excited-states ALMO-EDA formulation \cite{ge2018jcp,ge2018ene}, the dispersion energy cannot be readily separated from the frozen term. Zhao et al.\ \cite{zhao2023exc} have recently proposed a multistate EDA (MS-EDA) scheme formulated within a multistate DFT framework \cite{lu2022mul,lu2022fun} In this approach, the excimer interaction is represented by the exciton-excitation energy, superexchange (interfragment charge transfer), and an orbital-and-configuration delocalization contributions, the latter accounting for the energy lowering with respect to block-localized \cite{mo2000ene} wave function representatin of the complex. Complementary to EDA, widely adopted approaches analyze the character of the excimer wave function rather than the physical origin of the interaction. Such methods typically use one-particle density matrices \cite{luzanov2012exc,plasser2012ana,plasser2014new,herbert2024vis,krieger2025rat} or cumulants of two-particle density matrices \cite{luzanov2015qua}.

The paper is organized as follows. In Section 2, we briefly recall general equations for first-order degenerate Rayleigh-Schr{\"o}dinger perturbation theory (RSPT) \cite{hirschfelder1974deg}, discuss their application to intermolecular interactions, and extend them to the SAPT framework. Section 3 summarizes the computational details. In Section 4, we present numerical results for complexes exhibiting spatial and resonance-type degeneracies. Finally, Section~5 concludes the work.

\section{Theory}

\subsection{Degenerate Rayleigh--Schrödinger Perturbation Theory}

Let $H_0$ be an unperturbed Hamiltonian with a degenerate eigenspace $\mathcal{M}^0_I$ associated with the eigenvalue $E_I^{(0)}$. For simplicity, we restrict the discussion to the case $\dim \mathcal{M}^0_I=2$. The extension to higher-dimensional degenerate subspaces is straightforward. Let $|\phi_I^1\rangle$ and $|\phi_I^2\rangle$ be orthonormal eigenstates spanning this subspace:
\begin{equation}
H_0 |\phi_I^J\rangle = E_I^{(0)} |\phi_I^J\rangle, \quad J=1,2.
\end{equation}
We define the projection operator $P_0$ onto $\mathcal{M}^0_I$ as
\begin{equation}
P_0 = |\phi_I^1\rangle \langle\phi_I^1| + |\phi_I^2\rangle \langle\phi_I^2|.
\end{equation}
We seek the eigenstates of the perturbed Hamiltonian $H(\lambda)=H_0+\lambda V$ that evolve from the degenerate subspace $\mathcal{M}^0_I$ upon introducing the perturbation $V$. Specifically, we seek the states $|\tilde{\Psi}_{I_1}(\lambda)\rangle$ and $|\tilde{\Psi}_{I_2}(\lambda)\rangle$ satisfying
\begin{equation}\label{HlamdaElambda}
H(\lambda) |\tilde{\Psi}_{I_J}(\lambda)\rangle = E_{I_J}(\lambda)\, |\tilde{\Psi}_{I_J}(\lambda)\rangle, \quad J=1,2,
\end{equation}
with the condition $\lim_{\lambda \to 0} E_{I_J}(\lambda)=E_I^{(0)}$. We proceed by expanding $|\tilde{\Psi}_{I_J}(\lambda)\rangle$ and $E_{I_J}(\lambda)$ in powers of $\lambda$:
\begin{equation}\label{expansion_in_lambda}
\begin{split}
|\tilde{\Psi}_{I_J}(\lambda)\rangle &= |\tilde{\Psi}_{I_J}^{(0)}\rangle + \lambda |\tilde{\Psi}_{I_J}^{(1)}\rangle + \lambda^2 |\tilde{\Psi}_{I_J}^{(2)}\rangle+\ldots,
\\
E_{I_J}(\lambda) &= E_{I}^{(0)} + \lambda E_{I_J}^{(1)} + \lambda^2 E_{I_J}^{(2)}+\ldots.
\end{split}
\end{equation}
Using intermediate normalization, we impose
\begin{equation}
\langle \tilde{\Psi}^{(0)}_{I_J}|\tilde{\Psi}_{I_J}(\lambda)\rangle = \langle \tilde{\Psi}^{(0)}_{I_J}|\tilde{\Psi}^{(0)}_{I_J}\rangle = 1 \, ,
\end{equation}
We further assume that the perturbation lifts the degeneracy, so that
\begin{equation}
\langle \tilde{\Psi}_{I_1}(\lambda)|\tilde{\Psi}_{I_2}(\lambda)\rangle = 0 \, .
\end{equation}
The projector operator $P_0$ can be expressed as
\begin{equation}
P_0 = |\tilde{\Psi}_{I_1}^{(0)}\rangle \langle\tilde{\Psi}_{I_1}^{(0)}| + |\tilde{\Psi}_{I_2}^{(0)}\rangle \langle\tilde{\Psi}_{I_2}^{(0)}|.
\end{equation}
Substituting Eq.~\eqref{expansion_in_lambda} into Eq.~\eqref{HlamdaElambda} and collecting terms of equal order in $\lambda$ yields through the first order 
\begin{align}
H_0|\tilde{\Psi}_{I_J}^{(0)}\rangle &= E_I^{(0)}|\tilde{\Psi}_{I_J}^{(0)}\rangle,
\\ \label{lambda_expansion}
(V-E^{(1)}_{I_J})|\tilde{\Psi}_{I_J}^{(0)}\rangle &= (E_I^{(0)}-H_0)|\tilde{\Psi}_{I_J}^{(1)}\rangle.
\end{align}
Projecting Eq.~\eqref{lambda_expansion} onto $\mathcal{M}^0_I$ gives
\begin{equation}\label{pvp}
P_0VP_0|\tilde{\Psi}_{I_J}^{(0)}\rangle = E^{(1)}_{I_J}  P_0|\tilde{\Psi}_{I_J}^{(0)}\rangle.
\end{equation}
Thus, the zeroth-order states $|\tilde{\Psi}_{I_J}^{(0)}\rangle$ are eigenstates of the projected perturbation operator $P_0VP_0$, and the corresponding eigenvalues are the first-order energy corrections.

By expanding the zeroth-order wave function in the basis ${|\phi^1_I\rangle,|\phi^2_I\rangle}$,
\begin{equation} \label{psi0IJ}
|\tilde{\Psi}^{(0)}_{I_J}\rangle = c_{J,1}|\phi_I^1\rangle + c_{J,2}|\phi_I^2\rangle,
\end{equation}
we can rewrite Eq.~\eqref{pvp} as the matrix eigenvalue problem
\begin{equation}\label{matrixElstNEW}
\begin{bmatrix}
\color{white}\dot{\color{black}\bra{\phi_I^1}V\ket{\phi_I^1}} & \color{white}\dot{\color{black}\bra{\phi_I^1}V\ket{\phi_I^2}} \\[1ex]
\color{white}\dot{\color{black}\bra{\phi_I^2}V\ket{\phi_I^1}} & \color{white}\dot{\color{black}\bra{\phi_I^2}V\ket{\phi_I^2}}
\end{bmatrix}\begin{bmatrix}
c_{J,1} \\[1ex]
c_{J,2}
\end{bmatrix}=E_{I_J}^{(1)}\begin{bmatrix}
c_{J,1} \\[1ex]
c_{J,2}
\end{bmatrix} .
\end{equation}
Equation~\eqref{matrixElstNEW} provides a practical route for obtaining both the first-order energy corrections $E_{I_J}^{(1)}$ and the corresponding zeroth-order states through the coefficients $c_{J,1}$ and $c_{J,2}$. Assuming that the zeroth-order states $|\tilde{\Psi}^{(0)}_{I_J}\rangle$ are normalized such that $c_{J,1}^2+c_{J,2}^2=1$ and $E_{I_1}^{(1)}<E_{I_2}^{(1)}$, we can parametrize them using a mixing (diabatic) angle $\gamma_{\text{dRS}}\in[0,\pi)$, where
\begin{equation}
\begin{bmatrix}
c_{1,1} & c_{2,1} \\[1ex]
c_{1,2} & c_{2,2}
\end{bmatrix}=\begin{bmatrix}
\cos(\gamma_{\text{dRS}}) & -\sin(\gamma_{\text{dRS}}) \\[1ex]
\sin(\gamma_{\text{dRS}}) & \cos(\gamma_{\text{dRS}}).
\end{bmatrix}  
\end{equation}
Consequently, the mixing angle is computed as
\begin{equation}
  \tan(2\gamma_{\text{dRS}}) =  \frac{2 \braket{\phi_I^1|V|\phi_I^2} }{ \braket{\phi_I^1|V|\phi_I^1} - \braket{\phi_I^2|V|\phi_I^2} } \, .
\end{equation}
The formula resembles a single-property-based diabatization scheme, such as the Werner-Meyer dipole method \cite{werner1981mcs} or Boys localization \cite{subotnik2008con}, but relies on the interaction-matrix elements in the diabatic basis rather than molecular properties.

\subsection{Degenerate RSPT: Polarization Approximation}

To specialize the RSPT formalism to intermolecular interactions, the zeroth-order Hamiltonian is taken as the sum of the Hamiltonians of two noninteracting monomers,
\begin{equation}
H_0=H_A+H_B.
\end{equation}
The perturbation $V$ is identified with the intermolecular interaction operator
\begin{equation}
V = V_{AB} + \sum_{p\in A}v_B(p)+ \sum_{q\in B}v_A(q) + \sum_{\substack{p \in A \\ q \in B}}v_{ee}(p,q) \, ,
\end{equation}
where $V_{AB}$ denotes the nucleus--nucleus repulsion between the fragments, $v_B(p)$ and $v_A(q)$ describe the interaction of an electron in one monomer with the nuclei of the other monomer, and $v_{ee}(p,q)$ is the electron--electron interaction operator. We denote by $\ket{X_{i_j}}$ the $j$-th eigenstate of the monomer Hamiltonian $H_X$ with energy $E_i^X$:
\begin{equation}
H_X\ket{X_{i_j}} = E_{i}^X \ket{X_{i_j}}, \hspace{10mm}\text{for } X=A,B.
\end{equation}
The dimension of the eigenspace associated with the energy $E_{i}^X$ is denoted by $N^X_i$, so that $j=1,2,\ldots, N_{i}^X$. Whenever there is no degeneracy, i.e., $N_i^X=1$, the subscript $j$ will be omitted.

The eigenstates of $H_0$ can be represented as products of monomer eigenstates:
\begin{equation}
H_0|A_{k_l} B_{m_n}\rangle = E_I^{(0)} |A_{k_l} B_{m_n}\rangle,
\end{equation}
where $E^{(0)}_I=E_k^A+E_m^B$. The dimension of the eigenspace associated with the energy $E^{(0)}_I$ is at least $N^A_k N^B_m$. It is exactly $N^A_k N^B_m$ when monomers $A$ and $B$ are different molecules.

The degeneracy of the unperturbed solution may arise either from the spin degeneracy of the dimer or from the electronic degeneracy of one or both monomers. The problem of spin degeneracy has been addressed previously within the spin-flip SAPT framework developed by Patkowski and co-workers\cite{patkowski2018fir,waldrop2019spi}. In the present work, we focus on two representative cases of electronic degeneracy:
\begin{enumerate}
\item Spatial degeneracy. We consider monomer $A$ in a doubly-degenerate state ($N_0^A=2$) interacting with a nondegenerate monomer B ($N_0^B=1$). The zeroth-order degenerate subspace is spanned by two states
\begin{equation}
\begin{split}
\ket{\phi_0^1} &= |A_{0_1}B_0\rangle, \\
\ket{\phi_0^2} &= |A_{0_2}B_0\rangle, 
\end{split}
\end{equation}
where the superscript of $B_0$ has been omitted due to the absence of degeneracy. In the next section, we will refer to F(${}^2$P)$\cdots$H$_2$ and NO(${}^2\Pi$)$\cdots$H$_2$ as representative examples of spatial degeneracy.

\item Excimer degeneracy. We consider an excimer: two identical monomers in different electronic states, $k \neq m$. The energy of the general solution to the unperturbed problem then becomes
\begin{equation}
E_I^{(0)} = E^A_k + E^B_m = E^A_m + E^B_k,
\end{equation}
and the dimension of the corresponding eigenspace is equal to $2 N_k^A N_m^B$. Here, we assume a typical case in which the degeneracy arises solely from the indistinguishability of the monomers, that is, neither the ground state nor the excited state is itself degenerate ($N_{0}^A=N_{0}^B=1$ and $N_{\kappa}^A=N_{\kappa}^B=1$, where $\kappa$ denotes an excited state.) The zeroth-order solution is a linear combination of the two exciton-localized product states
\begin{equation}
\begin{split}
\ket{\phi_I^1} &= |A_\kappa B_0\rangle, \\
\ket{\phi_I^2} &= |A_0 B_\kappa\rangle.
\end{split}
\end{equation}
Because the zeroth-order model space is restricted to locally excited states, the present treatment describes excitonic resonance (Frenkel resonance) only and does not include the charge-transfer resonance mechanism.
\end{enumerate}

In both cases of electronic degeneracy, the diagonal matrix elements of $P_0VP_0$ in Eq.~\eqref{matrixElstNEW} correspond to the conventional first-order SAPT electrostatic energy, whereas the off-diagonal matrix elements describe the coupling between degenerate states. In the polarization approximation, these couplings are of purely electrostatic origin and can be expressed in terms of ground-state densities and transition densities. For spatial degeneracy, the coupling element takes the form
\begin{equation} \label{cplspa}
\begin{split}
\langle A_{0_1}B_0 | V | A_{0_2}B_0 \rangle &= \int \mathrm{d}1\, v_B(1)\rho_A^{0_1,0_2}(1) \\ 
&+ \int \mathrm{d}1\mathrm{d}2 \, \rho_A^{0_1,0_2}(1) v_{ee}(1,2) \rho_B^0(2),
\end{split}
\end{equation}
where $1=(\mathbf{r}_1,\sigma_1)$ denotes a combined spatial-spin coordinate. For excimer degeneracy, the coupling reads
\begin{equation}\label{ABvVAvB}
\langle A_0B_\kappa | V | A_\kappa B_0 \rangle =
\int \mathrm{d}1\mathrm{d}2\, \rho_A^{0,\kappa}(1) v_{ee}(1,2)\rho_B^{0,\kappa}(2).
\end{equation}
The electron density is defined as the diagonal part of the one-particle reduced density matrix (1-RDM)
\begin{equation}
\rho_X^{\mu_\nu}(1)=\gamma_X^{\mu_\nu}(1,1),
\end{equation}
where the 1-RDM is given by
\begin{equation}
\begin{split}
\gamma^{\mu_\nu}_{X}(1,1') =  N_X\int & \phi_{X}^{\mu_\nu}(1,2,\ldots,N_X) \\ 
\times & \phi_{X}^{\mu_\nu}(1',2,\ldots,N_X) \, \mathrm{d}2\ldots \mathrm{d}N_X \, .
\end{split}
\end{equation}
Here, $N_X$ is the number of electrons in monomer $X$ and $\phi_{X}^{\mu_\nu}(1',2,\ldots,N_X)$ is the spin-spatial representation of the $\ket{X_{\mu_\nu}}$ state. Similarly, the transition density is defined as the diagonal part of the one-particle transition reduced density matrix (1-TRDM),
\begin{equation}
\rho_X^{\mu_\nu,\kappa_\lambda}(1) = \gamma_X^{\mu_\nu, \kappa_\lambda}(1,1),
\end{equation}
where the 1-TRDM is given by
\begin{equation}
\begin{split}
\gamma_X^{\mu_\nu, \kappa_\lambda}(1,1') = N_X \int & \phi_X^{\kappa_\lambda}(1,2,\ldots,N_X) \\ 
\times & \phi_X^{\mu_\nu}(1',2,\ldots,N_X) \, \mathrm{d}2\ldots \mathrm{d}N_X.
\end{split}
\end{equation}
Note that, $\rho_X^{\mu_\nu,\kappa_\lambda}(1)=\rho_X^{\kappa_\lambda,\mu_\nu}(1)$, which was used in Eq.~\eqref{ABvVAvB}. Diagonalization of the matrix in Eq.~\eqref{matrixElstNEW} yields the first-order degenerate SAPT electrostatic energies, $E^{(1)}_{I_1,\mathrm{elst}}$ and $E^{(1)}_{I_2,\mathrm{elst}}$, together with the corresponding expansion coefficients of the zeroth-order wave functions.

The first term of the multipole expansion of Eq.~\eqref{ABvVAvB} is the well-known transition-dipole coupling (or the \textit{resonance dipole contribution}) responsible for excitonic splitting \cite{stone2013theory}:
\begin{equation}\label{StoneDipol}
\begin{split}
\langle A_0B_\kappa | V | A_\kappa B_0 \rangle & \approx \frac{(\mu_A)^{0\kappa}\cdot(\mu_B)^{0\kappa}}{R^3} \\
&- \frac{3[(\mu_A)^{0\kappa} \cdot \mathbf{R}][(\mu_B)^{0\kappa} \cdot \mathbf{R}]}{R^5},
\end{split}
\end{equation}
where $(\mu_X)^{0\kappa}$ denotes the transition dipole moment of monomer $X$ between states $|X_0\rangle$ and $|X_\kappa\rangle$. Thus, even if the molecules are nonpolar in both their ground and excited states, excitation can give rise to an interaction that decays asymptotically as $R^{-3}$.

\subsection{Degenerate Symmetrized RSPT}

Recall that the interaction energy expression in nondegenerate symmetrized Rayleigh–Schrödinger pertubation theory takes the form \cite{jeziorski1994per}
\begin{equation}\label{SRSenergy}
\mathcal{E}_I(\lambda)=\frac{\langle\tilde{\Psi}_I^{(0)}|\lambda V\mathcal{A}\tilde{\Psi}_I(\lambda)\rangle}{\langle\tilde{\Psi}_I^{(0)}|\mathcal{A}\tilde{\Psi}_I(\lambda)\rangle},
\end{equation}
where $\mathcal{A}$ denotes the antisymmetrizer, and $\ket{\tilde{\Psi}_I(\lambda)}$ follows from the polarization expansion of the wave function in the $I$-th state. The first-order exchange $E^{(1)}_{I,\text{exch}}$ energy is defined as the difference between the first-order energy obtained from Eq.~\eqref{SRSenergy} and the electrostatic energy $E^{(1)}_{I,\text{elst}}$. In the single-exchange approximation, electron exchange between the monomers is restricted to permutations involving a single pairs of electrons by truncating the antisymmetrizer as
\begin{equation}
    \mathcal{A}\approx\frac{N_A!N_B!}{(N_A+N_B)!}\, \mathcal{A}_A\mathcal{A}_B \,(1+\mathcal{P}_2),
\end{equation}
where $\mathcal{P}_2$ is the sum of all transpositions $\mathcal{P}_{pq}$ that interchange the coordinates of electrons $p$ belonging to monomer $A$ with those of an electron $q$  belonging to monomer $B$, $\mathcal{P}_2=-\sum_{p\in A}\sum_{q\in B}\mathcal{P}_{pq}$. In this framework, one neglects terms beyond
second order in the orbital overlap matrix $S$ and the first-order exchange energy can be obtained from one- and two-electron density matrices only.  
The SRS expression for the first-order exchange energy reads
\begin{equation}\label{exchs2SRS}
    E_{I,\text{exch}}^{(1)}=\langle \tilde{\Psi}_I^{(0)} \left|V\mathcal{P}_2\right| \tilde{\Psi}_I^{(0)}\rangle-\langle\tilde{\Psi}_I^{(0)}|V|\tilde{\Psi}_I^{(0)}\rangle\langle\tilde{\Psi}_I^{(0)}|\mathcal{P}_2|\tilde{\Psi}_I^{(0)}\rangle.
\end{equation}
This expression is simply the second cumulant of the $V$ and $\mathcal{P}_2$ operators. Thus, within the single-exchange approximation, Pauli repulsion may be viewed as a measure of the correlation between fluctuations (devations from the respective expectation values) of $V$ and $\mathcal{P}_2$.

In degenerate SRS theory, one may resolve the degeneracy in two distinct ways, which we refer to as antisymmetrization \textit{after} and \textit{before} the diagonalization. 
\begin{enumerate}
\item In the case of antisymmetrization \textit{after} diagonalization, we first diagonalize the $P_0VP_0$ operator to obtain the zeroth-order solution, $|\tilde{\Psi}_{I_J}^{(0)}\rangle$, cf.\ Eq.~\eqref{psi0IJ}. Next, we employ this solution in the standard SRS exchange energy expression, arriving at the formula identical to Eq.~\eqref{exchs2SRS}. However, the internal structure of Eq.~\eqref{exchs2SRS} is more complex compared to the regular nondegenerate SRS case, since in dSRS it requires access not only to 2-body reduced density matrixes (2-RDMs), but also 2-body transition reduced density matrixes (2-TRDMs). 
\item In antisymetrization \textit{before} diagonalization, we first include the antisymmetrizer already in the zeroth-order solution, 
\begin{equation}\label{PsiIj0}
\ket{\Psi^{(0)}_{I_J}} = a_{J,1} \mathcal{A}\ket{\phi_I^1} +a_{J,2}\mathcal{A}\ket{\phi_I^2} \, ,
\end{equation}
and only then solve the eigenproblem by projecting it at $\mathcal{M}^0_I$, to obtain:
\begin{equation}\label{ExchEigenproblemFull}
\begin{split}
& \begin{bmatrix}
\color{white}\dot{\color{black}\bra{\phi^1_I}V\mathcal{A}\ket{\phi^1_I}} & \color{white}\dot{\color{black}\bra{\phi^1_I}V\mathcal{A}\ket{\phi^2_I}} \\[1ex]
\color{white}\dot{\color{black}\bra{\phi^2_I}V\mathcal{A}\ket{\phi^1_I}} & \color{white}\dot{\color{black}\bra{\phi^2_I}V\mathcal{A}\ket{\phi^2_I}}
\end{bmatrix}\begin{bmatrix}
a_{J,1} \\[1ex]
a_{J,2}
\end{bmatrix} \\
&= \mathcal{E}_{I_J}^{(1)}
\begin{bmatrix}
\color{white}\dot{\color{black}\bra{\phi^1_I}\mathcal{A}\ket{\phi^1_I}} & \color{white}\dot{\color{black}\bra{\phi^1_I}\mathcal{A}\ket{\phi^2_I}} \\[1ex]
\color{white}\dot{\color{black}\bra{\phi^2_I}\mathcal{A}\ket{\phi^1_I}} & \color{white}\dot{\color{black}\bra{\phi^2_I}\mathcal{A}\ket{\phi^2_I}}
\end{bmatrix}
\begin{bmatrix}
a_{J,1} \\[1ex]
a_{J,2}
\end{bmatrix} .
\end{split}
\end{equation}
In the single-exchange approximation, Eq.~\eqref{ExchEigenproblemFull} takes the form:
\begin{equation}\label{ExchEigenproblem}
\begin{split}
& \begin{bmatrix}
\bra{\phi_I^1}V+V\mathcal{P}_2\ket{\phi^1_I} &
\bra{\phi^1_I}V+V\mathcal{P}_2\ket{\phi^2_I} \\[1ex]
\bra{\phi^2_I}V+V\mathcal{P}_2\ket{\phi^1_I} &
\bra{\phi^2_I}V+V\mathcal{P}_2\ket{\phi^2_I}
\end{bmatrix}
\begin{bmatrix}
a_{J,1}\\[1ex]
a_{J,2}
\end{bmatrix} \\
& =
\mathcal{E}_{I_J}^{(1)}
\begin{bmatrix}
1+\bra{\phi_I^1}\mathcal{P}_2\ket{\phi_I^1} &
\bra{\phi_I^1}\mathcal{P}_2\ket{\phi_I^2} \\[1ex]
\bra{\phi_I^2}\mathcal{P}_2\ket{\phi_I^1} &
1+\bra{\phi_I^2}\mathcal{P}_2\ket{\phi_I^2}
\end{bmatrix}
\begin{bmatrix}
a_{J,1}\\[1ex]
a_{J,2}
\end{bmatrix}.
\end{split}
\end{equation}
This way of resolving the degeneracy was first pursued by van Hemert and van der Avoird \cite{vanHemert1979abi}, and later by Chałasiński and Szalewicz \cite{chalbie1980deg}. The spin-summed density matrix expressions for the nondiagonal terms are given in the Supporting Information.
\end{enumerate}

Let us briefly compare the two approaches. When antisymmetrization is applied \textit{after} diagonalization, the degeneracy is resolved only through the electrostatic interaction by solving Eq.~\eqref{matrixElstNEW}. Consequently, the zeroth-order wave function, $\ket{\tilde{\Psi}^{(0)}_{I_J}}$ is identical as in the polarization approximation. In contrast, applying antisymmetrization \textit{before} diagonalization imposes the permutation symmetry already at the zeroth-order level, leading to $\ket{\Psi^{(0)}_{I_J}}$. Accordingly, the coefficients $c_{J,K}$ and  $a_{J,K}$, determined from Eqs.~\eqref{matrixElstNEW} and~\eqref{ExchEigenproblemFull}, respectively, are in general different.

Another important distinction concerns the decomposition of the first-order interaction energy. Antisymmetrization \textit{after} diagonalization preserves its separation into electrostatic and exchange contributions, 
\begin{equation}\label{E1int}
    E^{(1)}_{I_J,\text{int}}=E_{I_J, \text{elst}}^{(1)} + E_{I_J,\text{exch}}^{(1)},
\end{equation}
where $E_{I_J, \text{elst}}^{(1)}$  is obtained from  Eq.~\eqref{matrixElstNEW} and $E_{I_J,\text{exch}}^{(1)}$ follows from \eqref{exchs2SRS}. Conversely, antisymmetrization \textit{before} diagonalization entangles the electrostatic and exchange effects. As a result, the first order contribution, $\mathcal{E}^{(1)}_{I_J}$, can no longer be unambiguously divided into separate electrostatic and exchange terms.

The first-order energies obtained in the approaches are numerically different. While  $E^{(1)}_{I_J,\mathrm{int}}$ contains only terms through second order in $S$, $\mathcal{E}^{(1)}_{I_J}$ includes contributions of arbitrary order in the intermolecular overlap matrix, $S$: even though the antisymmetrizer is truncated to $S^2$ terms, diagonalization of Eq.~\eqref{ExchEigenproblem} yields disconnected terms of higher order in $S$. This holds even in the special case, when  $|\phi^1_I\rangle$ and $|\phi^2_I\rangle$ belong to different irreducible representations. Then, the off-diagonal terms in Eq.~\eqref{ExchEigenproblem} vanish and the solution takes the form  $ \mathcal{E}_{I_J}^{(1)} =\frac{\langle \phi^J_I| V+V\mathcal{P}_2|\phi^J_I\rangle} {\langle\phi^J_I|1+\mathcal{P}_2|\phi^J_I\rangle}$. It is easy to see that expanding the denominator generates an infinite series containing all even powers of the overlap matrix $S$. (This is the same reason why the $S^2$ approximation to the exchange energy is not exact even for the H$\cdots$H interaction.) 

Interestingly, it can be shown that $\mathcal{E}^{(1)}_{I_J}$ and $E^{(1)}_{I_J,\mathrm{int}}$ are related perturbatively. If the exchange interactions are much smaller than the electrostatic interactions, the exchange contribution can be treated as a perturbation to the electrostatic problem. In this case, $E^{(1)}_{I_J,\mathrm{int}}$ represents the first-order approximation to $\mathcal{E}^{(1)}_{I_J}$ (for details, see the Supporting Information).

Finally, note that antisymmetrization \textit{before} diagonalization leads to a non-hermitian eigenproblem: the $P_0 V\mathcal{A}P_0$ matrix on the left-hand side of Eq.~\eqref{ExchEigenproblem} is not hermitian, so that its eigenstates are no longer orthogonal. In principle, one should introduce two different mixing angles $\gamma_{\text{dSRS},1}$ and $\gamma_{\text{dSRS},2}$, for which we have that:
\begin{equation}\label{dSRS12}
\begin{bmatrix}
a_{1,1} & a_{2,1} \\[1ex]
a_{1,2} & a_{2,2}
\end{bmatrix}=\begin{bmatrix}
\cos(\gamma_{\text{dSRS},1}) & -\sin(\gamma_{\text{dSRS},2}) \\[1ex]
\sin(\gamma_{\text{dSRS},1}) & \cos(\gamma_{\text{dSRS},2})
\end{bmatrix} . 
\end{equation}
We verified that in all the studied cases the differences between $\gamma_{\text{dSRS},1}$ and $\gamma_{\text{dSRS},2}$ are less than $5^\circ$. Thus, we put $\gamma_{\rm dSRS}=\frac{1}{2}(\gamma_{\text{dSRS},1}+\gamma_{\text{dSRS},2})$.

The presented formalism assumes that the degeneracy is lifted at first order. If this is not the case, the effective interaction must be constructed through second order in $\lambda$ and diagonalized within the degenerate manifold. Although a detailed discussion of this problem is deferred to future work, we note that the resulting eigenvectors reproduce the $a_{J,K}$ coefficients, and hence the mixing angle, through first order in $\lambda$.

Following the convention commonly adopted in the literature, we henceforth use the term degenerate SAPT (dSAPT) to refer to degenerate SRS theory. 

\section{Computational details}

Degenerate SAPT calculations based on CASSCF description on monomers were performed in the Gammcor code \cite{gammcor}. Degenerate SAPT calculations employing CIS monomer wave functions were carried out with a separate Psi4NumPy \cite{psi4numpy} implementation. The required one- and two-electron integrals and monomer CASSCF one-, and two-electrons density matrices were obtained either from Molpro \cite{molpro2012} or the Quantum Package \cite{qpackage2}. The interface between Quantum Package and GammCor was implemented using the TREXIO library \cite{posenitskiy2023tre}. For the dSAPT(CAS) calculations on NO$\cdots$H$_2$, the H$_2$O excimer, and the benzene excimer, approximate 2-TRDMs were constructed from zeroth-order extended RPA (ERPA) eigenvectors and 2-RDMs \cite{pernal2012exc,pernal2014how,hapka2019sec}. The target ERPA eigenstate was identified by comparing the ERPA eigenvectors with the corresponding CASSCF vectors reconstructed from 1-TRDM, and selecting the ERPA state with the largest overlap.\cite{drwal2021exc,drwal2024mul} The quality of the approximate 2-TRDMs was assessed in two ways. First, for the H$_2$O$\cdots$H$_2$O$^*$ excimer, we compared dSAPT results obtained with exact (CASSCF) and approximate (ERPA-based) 2-TRDMs, finding satisfactory agreement (see Figure~S1 in the Supporting Information). Second, we contracted the approximate 2-TRDMs to the corresponding 1-TRDMs and compared the result with the 1-TRDMs obtained directly from the CASSCF calculations. The latter comparison was also used to fix a consistent global phase of the 2-TRDMs.

Computational scaling of the first-order dSAPT(CAS) method is $n_{\rm occ}^6$, where $n_{\rm occ}$ represents the number of occupied (sum of inactive and active) orbitals. In dSAPT(CIS), we evaluate contractions directly using explicit form of the RDMs, which leads to the formal scaling of $n_{\rm occ}^3n_{\rm virt}^2$, where $n_{\rm virt}$ denotes the number of virtual orbitals. The density-fitted dSAPT(CIS) formulas were also implemented. In this case, the formal scaling is $n_{\rm occ}^2n_{\rm virt}N_{\rm aux}$, where $N_{\rm aux}$ denotes the size of auxiliary basis set. The dSAPT(CIS) implementation was cross-validated against automatically derived second-quantized formulas \cite{tyrcha2024sec}.

For nondegenerate SAPT, we represent the interaction energy through second order in $V$ as:
\begin{equation}
E_{\text{int}}^{\text{SAPT}} = E^{(1)}_{\rm elst} + E^{(1)}_{\rm exch} + E^{(2)}_{\rm ind} + E^{(2)}_{\rm exch-ind} + E^{(2)}_{\rm disp} + E^{(2)}_{\rm exch-disp} \, ,
\end{equation}
where $E^{(2)}_{\rm ind}$ and $E^{(2)}_{\rm disp}$ are induction and dispersion energies, respectively, accompanied by their exchange components. Second-order terms are obtained at the coupled level of theory. All exchange energy terms employ the $S^2$ approximation, unless stated otherwise.

In potential-energy scans, the signs of the off-diagonal matrix elements and of the corresponding mixing angles were adjusted to ensure smooth variation with the nuclear geometry. Such adjustment is necessary because the overall phase of the CI/CAS wave functions is arbitrary, which may otherwise lead to discontinuous sign changes in transition quantities between neighboring
geometries. In complexes with spatial degeneracy, monomer molecular orbitals for open-shell species (F and NO) were taken from reference geometries in the asymptotic region ($R=50~a_0$). In this way, we avoid dependence of the monomer molecular orbitals on the basis set of the partner, which is similar to the monomer-centered basis set approach \cite{williams1995ont}.

\section{Results}

\subsection{F($^2\mathrm{P}$)$\cdots$H$_2$($^1\Sigma_g^+$)}
The F$\cdots$H$_2$ dimer is a paradigm example of a system governed by multiple potential energy surfaces.
It has been studied extensively by both theory \cite{rebentrost1975non,alexander2000ani,aquilanti2001pot,klos2002abi,klos2004par} and experiment \cite{neumark1985mol,skodje2000res,qiu2006obs}, with accurate model potentials guiding state-of-the-art scattering measurements \cite{tizniti2014rate}. Here, we focus on three lowest states resulting from the coupling between ground-state H$_2$ and a triply degenerate fluorine $^2\mathrm{P}$ atom. Using dSAPT, we can verify how electrostatics and exchange contribute to the nonadiabatic coupling between interacting states.

The H$_2$ bond length is fixed at $r_{\mathrm{HH}} = 1.448\,a_0$. The geometry is described using Jacobi coordinates, where $R$ denotes the distance between the fluorine atom and the center of mass of H$_2$, while $\theta$ is the angle between the intermolecular vector $\mathbf{R}$ and the molecular axis $\mathbf{r}$. We define $\theta=0^\circ$ as the collinear geometry. The coordinate system is chosen such that the fluorine atom is located at the origin, the vector $\mathbf{R}$ is always aligned with the $z$ axis, and the $x$ axis lies in the plane containing all atoms.

\begin{figure}
\centering
\includegraphics[width=\linewidth]{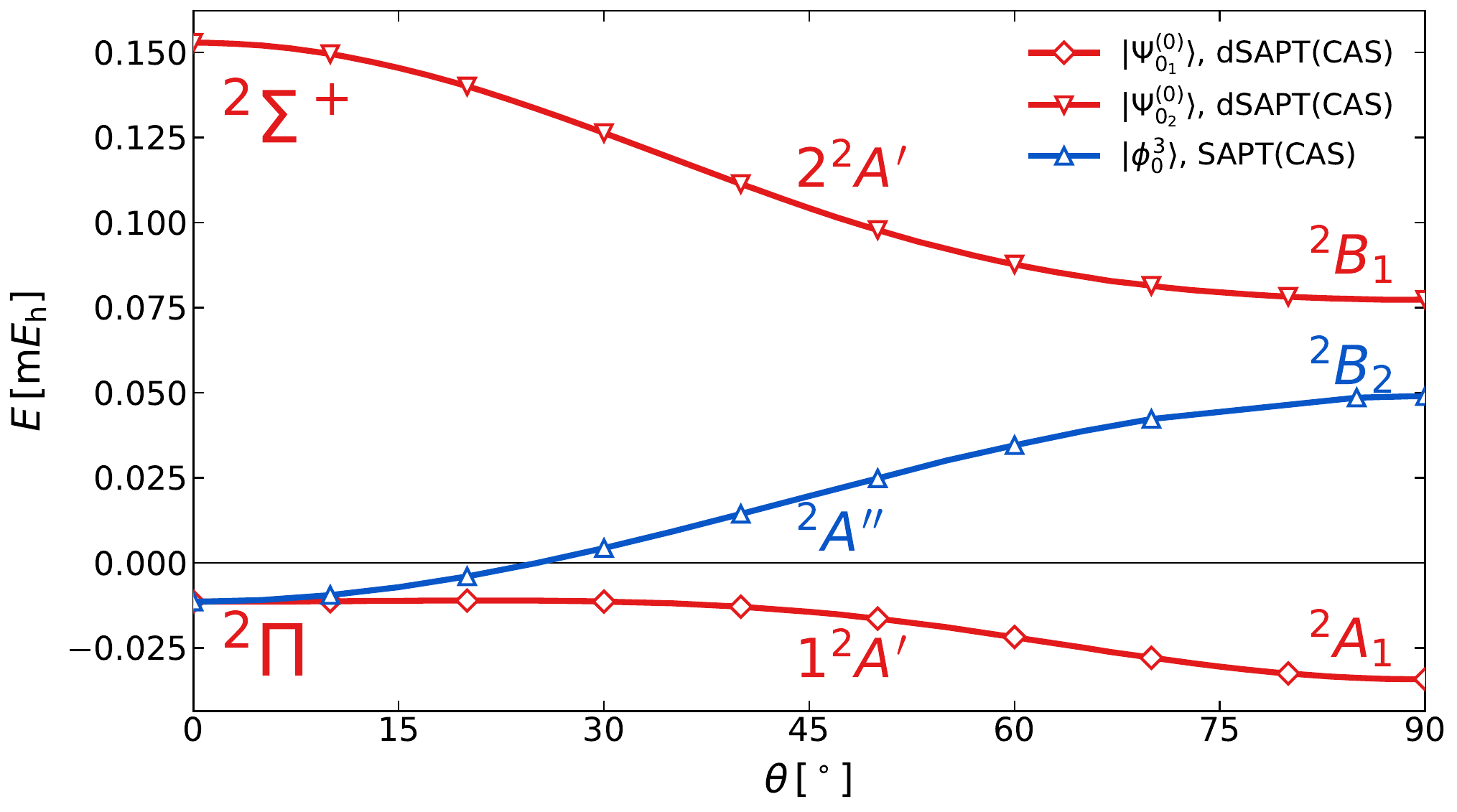}
\caption{Angular dependence of the first-order SAPT interaction energy for the F($^2$P)$\cdots$H$_2$ complex at $R=7.0\,a_0$, together with the corresponding state symmetries. The red curves were obtained by diagonalizing the interaction matrix after antisymmetrization, according to Eq.~\eqref{ExchEigenproblem}. The blue curve represents the first-order nondegenerate SAPT interaction energy calculated using the $|\phi_0^3\rangle$ wave function. \label{FH2_sym}}
\end{figure}

The orbital degeneracy of the fluorine atom in its ground ${}^2$P state gives rise to $^{2}\Sigma^+$ and $^{2}\Pi$ states in the linear $C_{\infty v}$ geometry. Upon lowering the symmetry to $C_s$ ($\theta\notin\{0^\circ,90^\circ\}$), these states transform into two states of $A'$ symmetry and one state of $A''$ symmetry. At the T-shaped $C_{2v}$ geometry, they decouple into the $^{2}A_1$, $^{2}B_1$, and $^{2}B_2$ states. In Figure~\ref{FH2_sym}, we illustrate the relation between the considered states by plotting the first-order SAPT adiabatic bending potentials. In these calculations, the F(${}^2$P) atom and the H$_2$ molecule are described at the CAS(5,3)SCF and Hartree-Fock levels, respectively.

In the $\theta\in(0^\circ,90^\circ)$ range, the $A''$ state corresponds to a singly occupied $p_y$ orbital on the fluorine atom. Because this state is symmetry-forbidden from coupling to the two $A'$ states, it can be treated using standard, nondegenerate SAPT.  In contrast, describing the interacting $A'$ states requires degenerate SAPT formulation. We define the zeroth-order degenerate space basis functions as:
\begin{equation}
|\phi_0^1\rangle = |A_{0_1}B_0\rangle, 
\qquad
|\phi_0^2\rangle = |A_{0_2}B_0\rangle,
\end{equation}
where $\ket{B_0}$ denotes the H$_2$ wave function, 
and $\ket{A_{0_1}}$ and $\ket{A_{0_2}}$ are the spatially degenerate 
atomic wave functions of the fluorine atom:
\begin{equation}
      |A_{0_1}\rangle = |1s^2 2s^2 2p_y^2 2p_x^2 2p_z^1 \rangle, \, \, |A_{0_2}\rangle = |1s^2 2s^2 2p_y^2 2p_x^1 2p_z^2 \rangle.
\end{equation}
Consequently, we denote the $A''$ state by $\ket{\phi_0^3}$.

\begin{figure*}
\centering
\includegraphics[width=1.0\linewidth]{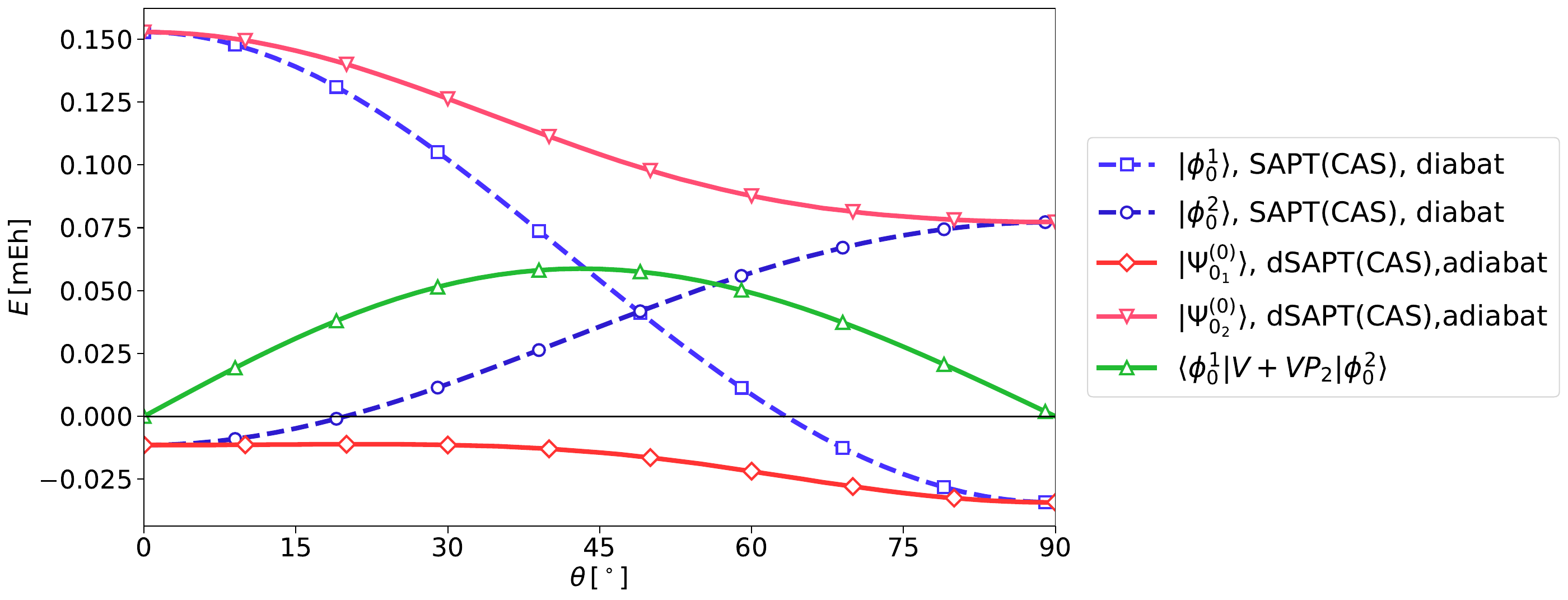}
\caption{Angular dependence of the first-order SAPT diabatic (blue dotted) and adiabatic (red solid) interaction energy surfaces for the F($^2$P)$\cdots$H$_2$ complex at $R=7.0\,a_0$. The green curve shows the off-diagonal coupling matrix element $\langle\phi^1_0|V+V\mathcal{P}_2|\phi^2_0\rangle$. The adiabatic surfaces are obtained by solving the generalized eigenvalue problem of Eq.~\eqref{ExchEigenproblem}, wheras the blue dotted lines are obtained by using first-order nondegenerate SAPT for states $|\phi_1\rangle$ and $|\phi_2\rangle$.}
\label{Matrix elements}
\end{figure*}

We begin by analyzing the role of the first-order exchange in dSAPT, using CASSCF description of the monomers [dSAPT(CAS)]. In Figure~\ref{Matrix elements}, we present the angular dependence of the first-order coupling element alongside the SAPT diabatic and adiabatic potential energy surfaces at $R = 7.0\,a_0$. At the SAPT(CAS) level, the global minimum occurs for the T-shaped ${}^2A_1$ geometry at approximately 5.8~$a_0$ (see Figure S2 in the Supporting Information). Thus, the distance $R = 7.0\,a_0$ corresponds to roughly $1.2 \, R_{\mathrm{eq}}$. At this separation, energy terms beyond the $S^2$ approximation are negligible, meaning that  antisymmetrization \textit{before} and \textit{after} diagonalization lead to practically identical results (see Figure S3 in the Supporting Information).

As is evident from Figure~\ref{Matrix elements}, the coupling reaches maximum at $\theta=45^{\circ}$, where it becomes comparable in magnitude to the SAPT(CAS) diabats. To partition the coupling into its electrostatic and exchange components, in Table~\ref{MatrixElementsdSRS} we compare the matrix elements of Eq.~\eqref{ExchEigenproblem} at $R = 7.0\,a_0$ and $\theta=45^{\circ}$. The last two columns reveal that the state mixing is dominated by the electrostatic term. The off-diagonal exchange contribution, $\langle\phi^1_0|V\mathcal{P}_2|\phi^2_0\rangle$, is small, accounting for less than 1$\%$ of the total coupling element. Nevertheless, the exchange energy remains essential, as it dramatically shifts the diagonal elements of $P_0V\mathcal{A}P_0$. Although the eigenproblem in Eq.~\eqref{ExchEigenproblem} is not hermitian, this violation is weak: the off-diagonal $V\mathcal{P}_2$ elements differ by only 0.06~$\mu E_h$.

\begin{table}
\centering
\caption{Comparison of dSAPT(CAS) matrix elements appearing in Eq.~\eqref{ExchEigenproblem} for the F$\cdots$H$_2$ dimer. The results correspond to $R=7.0\,a_0$, $\theta=45^\circ$. Expected values of $V$ and $V\mathcal{P}_2$ operators are given in $\mu E_h$. The  dimensionless $\mathcal{P}_2$ matrix elements are scaled by the factor of $10^{-6}$.}
\label{MatrixElementsdSRS}
\begin{tabular}{l r r r r} \hline
$\hat{O}$ & $\langle\phi_0^1|\hat{O}|\phi_0^1\rangle$
 & $\langle\phi_0^2|\hat{O}|\phi_0^2\rangle$
 & $\langle\phi_0^1|\hat{O}|\phi_0^2\rangle$
 & $\langle\phi_0^2|\hat{O}|\phi_0^1\rangle$ \\
\hline
$V$ &
$-19.4$ &
$18.7$ &
$59.1$ &
$59.1$ \\

$V\mathcal{P}_2$ &
$55.1$ &
$35.4$ &
$-0.53$ &
$-0.47$ \\

$\mathcal{P}_2$ &
$-177.1$ &
$-101.3$ &
$1.8$ &
$1.8$ \\
\hline
\end{tabular}
\end{table}

\begin{figure*}
\centering
\includegraphics[width=1.0\linewidth]{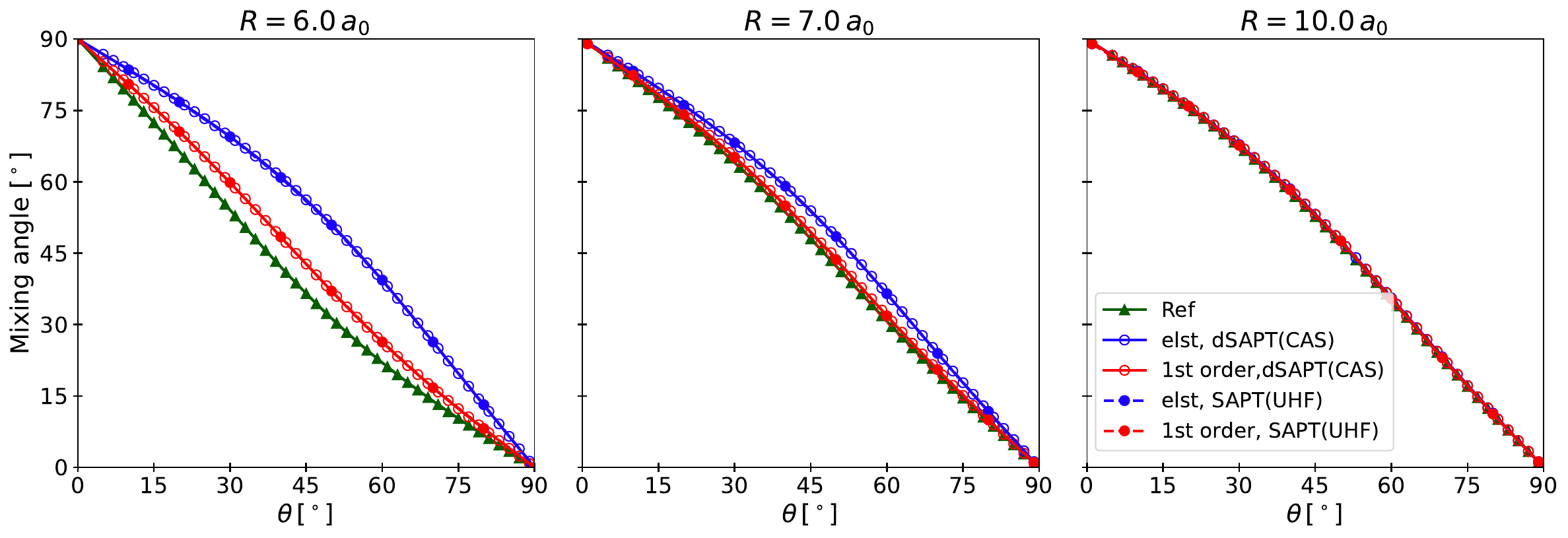}
\caption{Mixing angle as a function of $\theta$ for the F$\cdots$H$_2$ complex at $R=6.0$, $ 7.0$ and $10.0 \,a_{0}$. The reference values were obtained using the method of Alexander \cite{alexander1993adi}. The ``elst, dSAPT(CAS)'' and ``1st order, dSAPT(CAS)'' curves correspond to the solutions of Eqs.~\eqref{matrixElstNEW} and~\eqref{ExchEigenproblem}, respectively. The curves labeled ``elst, SAPT(UHF)'' and ``1st order, SAPT(UHF)'' were obtained using the nondegenerate SAPT(UHF) procedure described in Eq.~\eqref{SAPT(UHF)elst} and  Eq.~\eqref{SAPT(UHF)exch}, respectively.}
\label{FH2mixing}
\end{figure*}

The coupling between two interacting states of the same symmetry is conveniently characterized by the mixing angle. In Figure~\ref{FH2mixing}, we compare the mixing angles obtained from dSAPT with the reference supermolecular CASSCF values of Alexander \cite{alexander1993adi}, where the mixing angle is determined from transition matrix elements of the angular momentum operator. In dSAPT, the antisymmetrization \textit{after} diagonalization yields mixing angles that account only for the electrostatic coupling, the same as obtained within polarization approximation of Eq.~\eqref{matrixElstNEW}. Consequently, we label these results as ``elst, dSAPT(CAS)'' in Figure~\ref{FH2mixing}. Conversely, when antisymmetrization \textit{before} diagonalization is employed [Eq.~\eqref{ExchEigenproblem}], both electrostatic and exchange interactions contribute to the state mixing (in Figure~\ref{FH2mixing}), we use the ``1st order, dSAPT(CAS)'' label.

At large intermolecular separations ($R=10~a_0$ panel in Figure~\ref{FH2mixing}), the mixing angles obtained from  electrostatics and first-order dSAPT calculations are nearly identical, reflecting the negligible role of exchange effects in this regime.
As the fluorine atom approaches the H$_2$ molecule, exchange contributions become increasingly important. This trend is already visible at intermediate intermolecular separations ($R=7.0\,a_0$)  and becomes particularly pronounced near the equilibrium distance  ($R=6.0\,a_0$). Importantly, including the first-order exchange correction substantially reduces the discrepancy with the reference results.

The F($^2$P)$\cdots$H$_2$ dimer is a special case of spatial degeneracy where the mixing angle can be computed within the single-reference SAPT(UHF) \cite{Hapka:12} framework. In essence, an eigenstate of the (symmetry-adapted) perturbation operator can be found simply by rotating the 2$p$ orbitals in the fluorine UHF wave function. The positions of the extrema of the SAPT(UHF) energy as function of the orbital rotation angle then correspond to the mixing angles. The details of this procedure are provided in the Appendix.

In Figure~\ref{FH2mixing}, we demonstrate that the mixing angles obtained with the SAPT(UHF) method remain in perfect agreement with those from dSAPT(CAS), validating our dSAPT implementation. The mixing angles obtained from first-order SAPT calculations closely follow the supermolecular CAS(5,3)SCF reference at large separations, but visibly deviate in the van der Waals minimum region. Because the active orbitals are localized only on the fluorine atom, the supermolecular reference does not account for dispersion interactions \cite{Hapka:20}. Still, the inclusion of second-order induction contributions at the SAPT(UHF) level brings only a minor improvement (see Figure S4 in the Supporting Information). Thus, we attribute this discrepancy to higher-order effects.
     
For linear geometry at fixed H$_2$ distance $\Pi$ and $\Sigma$ state cross for $R=4.8$~$a_0$ and for small $\theta$ the crossing is replaced by an avoided crossing (on a 3D surface, the two surfaces intersect on a seam of linear geometry). dSAPT(CAS) predicts the position crossing quite well, at 5.4~$a_0$, given the lack of higher orders and intramonomer correlation in present theory.

\subsection{H$_2$($^1\Sigma_g^+$)$\cdots$NO($^2\Pi$)}

The second spatially degenerate complex we consider is the NO radical in the $^2\Pi$ state interacting with the ground-state H$_2$ molecule. Diabatization in molecule-molecule dimers is generally more challenging compared to the atom-diatom case due to the lack of planar symmetry. In the general, nonplanar arrangement, the two lowest electronic states of the H$_2\cdots$NO(${^2}\Pi$) dimer belong to the same irreducible representation and may thus interact.

The NO and H$_2$ bond lengths are set to $r_{\text{NO}} = 2.1803~a_0$ and $r_{\rm H_2} = 1.448~a_0$. With rigid monomers, the interaction energy is a function of four internal coordinates: the distance between centers-of-mass $R$ of the molecules, the $\theta_{\rm NO}$ and $\theta_{\rm H_2}$ polar angles, and the azimuthal angle $\phi_H$. The angles $\theta_{\rm NO}=0^\circ$ and $180^\circ$ correspond to the nitrogen and oxygen atoms, respectively, directed towards the center of the H-H bond (see Figure S5 for the choice of the coordinate system).

In contrast to the F$\cdots$H$_2$ interaction, the coupling via first-order exchange in H$_2\cdots$NO cannot be neglected. In particular, the off-diagonal matrix elements of $P_0V\mathcal{P}_2P_0$ are comparable in magnitude to electrostatic $\langle \phi_{0}^1|V|\phi_{0}^2\rangle$ contributions for nonplanar H-shaped geometries [$\theta_{\mathrm{NO}}\in(40^\circ,140^\circ)$, see Figure~S6 in the Supporting Information]. Moreover, the effect of the non-Hermiticity of $P_0V\mathcal{P}_2P_0$ is more pronounced compared to the F$\cdots$H$_2$ complex. This leads to differences between the mixing angles $\gamma_{\mathrm{dSRS},1}$ and $\gamma_{\mathrm{dSRS},2}$, defined in Eq.~\eqref{dSRS12}, of up to $5^\circ$ in the repulsive wall region. 

\begin{figure*}
\centering
\includegraphics[width=0.95\linewidth]{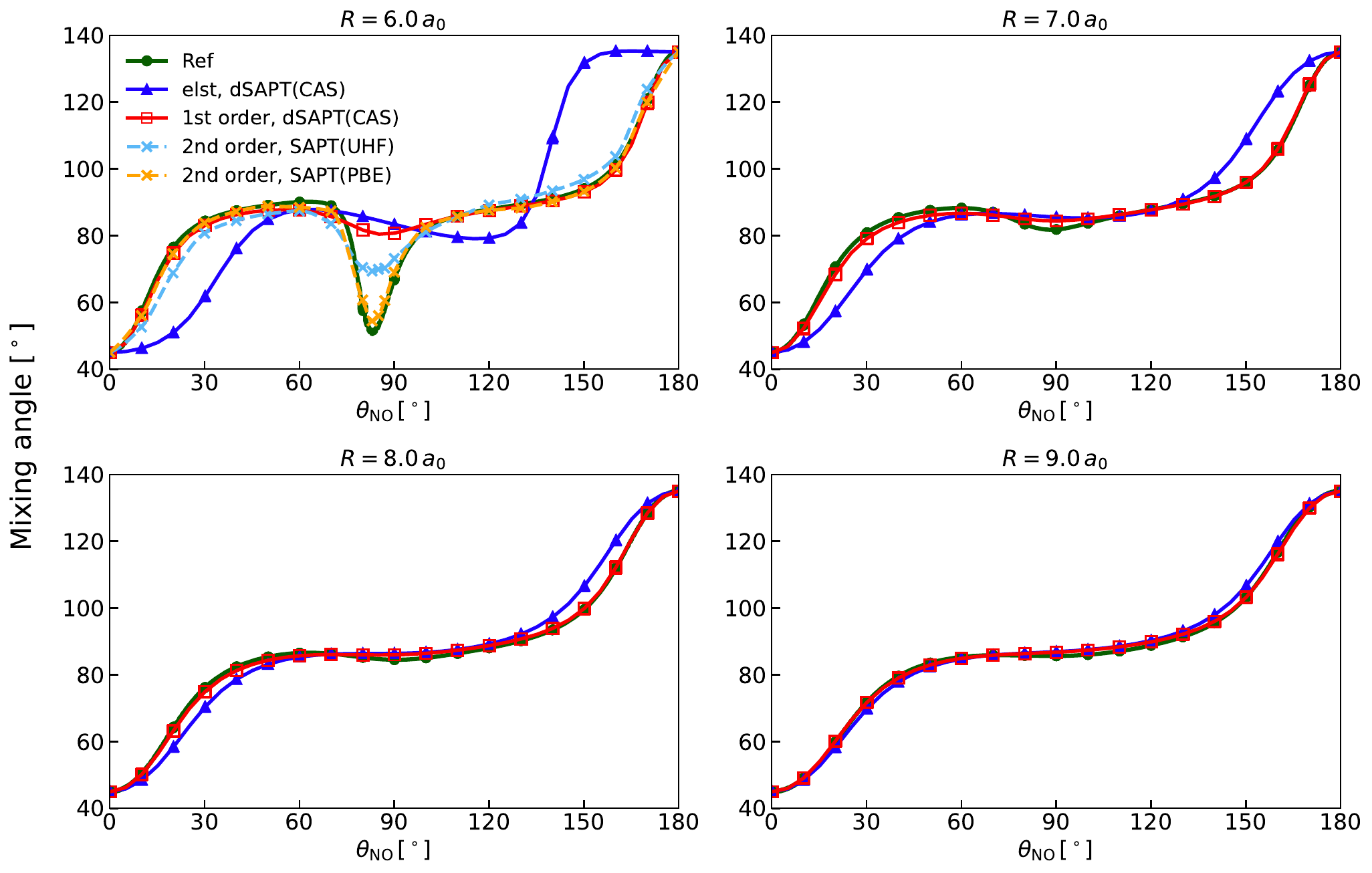}
\caption{Mixing angle as a function of $\theta_{\rm NO}$ for the H$_2\cdots$NO(${}^2\Pi$) complex at $R=6.0$, $7.0$, $8.0$, and $9.0~a_0$ at a nonplanar geometry defined by fixed $\theta_{\rm H_2}=90^\circ$ and $\phi=45^\circ$ angles. The ``elst, dSAPT(CAS)'' and ``1st order, dSAPT(CAS)'' curves correspond to the solutions of Eq.~\eqref{matrixElstNEW} and Eq.~\eqref{ExchEigenproblem}, respectively. The curves labeled ``2nd order, SAPT'' were obtained using the nondegenerate SAPT procedure described in the Appendix. Reference values taken from Ref.\citenum{deJongh2017ima}.}
\label{fig:noh2}
\end{figure*}

Figure~\ref{fig:noh2} displays the mixing angle as a function of the $\theta_{\text{NO}}$ angle for selected values of $R$ at nonlinear geometry defined by fixed Jacobi coordinates: $\theta_{\rm H_2} = 90^\circ$ and $\phi_H=45^\circ$. At the dSAPT(CAS) level, the mixing angles are computed either from purely electrostatic coupling [Eq.~\eqref{matrixElstNEW}] or the first-order coupling [Eq.~\eqref{ExchEigenproblem}]. The NO and H$_2$ molecules are described using CAS(7,6)SCF and CAS(2,2)SCF wave functions, respectively. Also reported are mixing angles from second-order nondegenerate SAPT calculations based on UHF and UKS description of the monomers (lines labeled as ``2nd order, SAPT'' in Figure~\ref{fig:noh2}), computed using the procedure described in the Appendix (in analogy to F$\cdots$H$_2$, the mixing angle is  obtained by rotating the $\pi^*$ orbitals in the unrestricted Hartree-Fock or Kohn-Sham determinant of the NO molecule). SAPT(UKS) calculations employed the PBE functional \cite{Perdew:96} asymptotically-corrected using the gradient-regulated connection scheme \cite{Gruning:01}. As reference, we adopted partially spin-restricted coupled-cluster PES with single and double excitations and perturbative triples [RCCSD(T)] of de Jongh et al. \cite{deJongh2017ima} computed using multiple-property-based diabatization algorithm of Karman et al. \cite{karman2016com}

At large intermolecular separations, configuration mixing is governed solely by the electrostatic quadrupole-quadrupole interaction \cite{wormer2005abi,karman2018dia}. The exchange term begins to contribute already at $R$=$9~a_0$, particularly for $\theta_{\text{NO}} \in (0^\circ,60^\circ)$ and $\theta_{\rm NO} \in (120^\circ,180^\circ)$, see Figure~\ref{fig:noh2}. This reflects the fact that Pauli repulsion is most pronounced in T-shaped geometries, when the overlap between NO and \ce{H2} is largest (see also Figure~S7 in the Supporting Materials). At $R$=$6~a_0$, the effect of exchange is strong for all NO orientations. This distance corresponds to the repulsive wall for T-shaped geometries, and probes the vicinity of the minimum in the skewed H-shaped arrangement (see Figures S8-S9 for radial cuts of the PES in high-symmetry geometries).

As evident from Figure~\ref{fig:noh2}, the electrostatic and exchange energy contributions alone do not reproduce the qualitative behavior of the mixing angle. At short intermolecular distances, a minimum appears near $\theta_{\rm NO}=85^\circ$ and this feature is recovered only after the second-order SAPT contributions are included. Quantitative agreement with the reference RCCSD(T) results is obtained when intramonomer correlation effects are accounted for at the SAPT(PBE) level. SAPT(UHF) correctly predicts the position of the mixing angle minimum, but underestimates its depth. A closer analysis reveals that the dispersion energy contributes substantially to the state mixing, while the minimum itself results from the interplay of electrostatic, exchange and dispersion components.

To explain how the mixing angle minimum forms, recall that the mixing angle can be determined by \textit{changes} in the SAPT energy components upon rotation of the $\text{NO}$ molecule's $\pi^*$ orbitals [see Eq.\eqref{SAPT(UHF)exch} in the Appendix]. At small and large $\theta_{\rm NO}$ angles, the electrostatic and exchange contributions shift in tandem under this rotation. Near $\theta_{\rm NO}=80^\circ$, however, the exchange term changes sign and counteracts electrostatics (see Figure S10 in the Supporting Information). In Figure~\ref{fig:noh2}, this competition is reflected in the relative shift between the electrostatic and first-order curves. Furthermore, in nonplanar H-shaped geometries, Pauli repulsion becomes relatively insensitive to NO orientation, leaving first-order exchange variations small. Because second-order induction and exchange-induction energies largely cancel out, the dispersion term emerges as the dominant driver of state mixing, shaping the pronounced minimum at short intermolecular distances.

\subsection{The H$_2$O$\cdots$H$_2$O$^*$ excimer}

First-order dSAPT serves as a rigorous perturbative framework for going beyond the electrostatic (Coulomb) coupling that governs the long-range degenerate interactions. At shorter intermolecular distances, the overlap between subsystem densities becomes non-negligible, and exchange interactions begin to influence state mixing. In excimers and excitons, exchange contributions can be approximated using the Dexter model \cite{dexter1953ath}, which assumes orthogonality between the subsystem wave functions. Nevertheless, intermonomer penetration terms are also relevant and should be taken into account \cite{scholes1995rat,harcourt1996ont,scholes2002ele,fuckel2008the,barcza2026ont}. Several extensions of the Frenkel-Davydov exciton model \cite{frenkel1931ont,davydov1964} have been developed to include both exact (Hartree-Fock) exchange and overlap effects \cite{yamagata2012des,zhang2012imp,ma2013cal,morrison2014abi,kaiser2023amu,pitesa2024exc}. In degenerate SAPT, these effects emerge naturally already at first order as a consequence of antisymmetrization.

The lowest singlet excited-state potential energy surface of the water dimer has been studied primarily in the context of hydrogen transfer along stretching curves of the hydrogen-bonded OH moiety \cite{sosa1993mul,kowal2001the,sobolewski2002sob,chipman2006str}. The first \textit{ab initio} investigation of the water excimer was, in fact, a degenerate SAPT study by van Hemert and van der Avoird \cite{vanHemert1979abi}, who attributed the solvation blue shift in the ${}^1A_1 \to {}^1B_1$ water band to increased exchange repulsion relative to the ground state. Here, we revisit this system and highlight the interplay of electrostatics and Pauli repulsion for the lowest-lying singlet $^{1}B_{1}( 1b_{1}\rightarrow 3s/4a_{1})$ excited state of the monomers. 

\begin{figure}
\centering
\includegraphics[width=1\linewidth, 
    width=\linewidth,
    trim=0 0 10cm 0,
    clip]{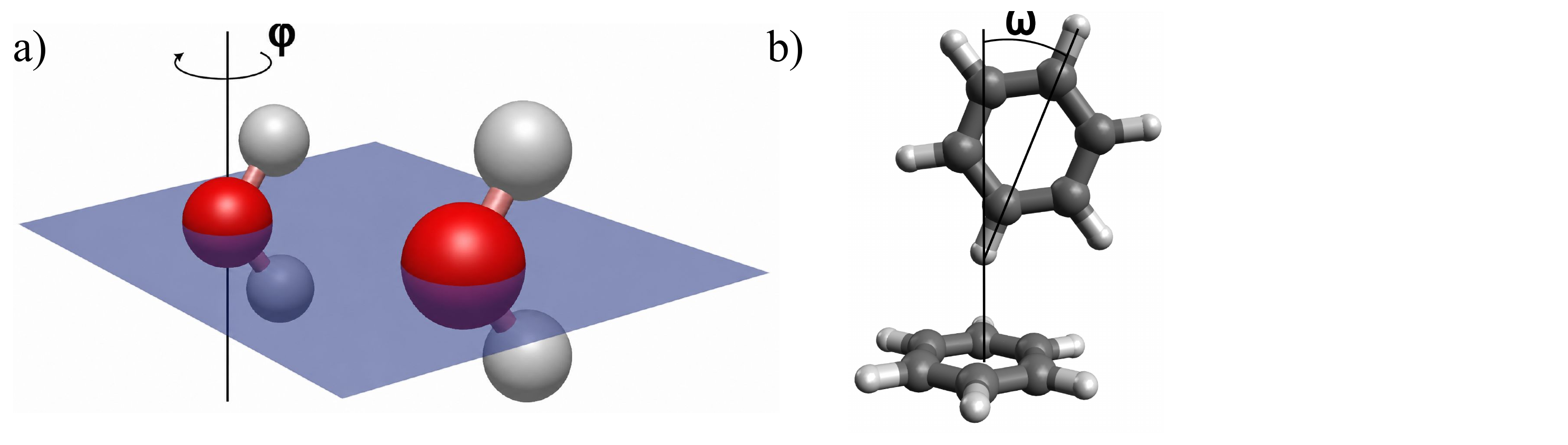}
\caption{Schematic representation of the model excimer geometries and definitions of the angular coordinates used in the scans: (a) the rotation angle $\varphi$ for the water excimer and (b) the tipping angle $\omega$ for the benzene excimer.}
\label{geoms}
\end{figure}

To probe the dependence of the excited-state interaction on the mutual orientation of the monomers, we employ a simple constrained model of the water dimer (see Figure~\ref{geoms}). The internal geometry of each water molecule is kept fixed throughout the calculations, with an O--H bond length of $r_{\mathrm{OH}}=0.9568~\text{\AA}$ and an H--O--H bond angle of $104.9^{\circ}$. The distance between the oxygen atoms is fixed at $R=7.0~a_{0}$, while the relative orientation of the two water molecules is described by the angle $\varphi$. The chosen intermolecular separation samples the vicinity of the ground-state minimum for $\varphi\in(90^\circ,180^\circ)$ and the repulsive wall for the remaining orientations. The reference geometry, corresponding to $\varphi=0^{\circ}$, consists of two parallel identically oriented water molecules and belongs to the $C_{2v}$ point group. The angle $\varphi$ is introduced by rigidly rotating one of the water molecules about an axis passing through its oxygen atom and perpendicular to the symmetry plane containing both oxygen atoms. The direction of rotation is defined such that, for $0^{\circ}<\varphi<180^{\circ}$, the hydrogen atoms of the rotating monomer are displaced toward the intermolecular region. At $\varphi=180^{\circ}$, the two water molecules recover a parallel but oppositely oriented arrangement, forming another high-symmetry configuration that belongs to the $C_{2h}$ point group.

The coupling through exchange plays an important role in the water excimer. This is illustrated in Table~\ref{MatrixElementsdSRSH2O} which reports the relevant matrix elements for a representative $C_{2v}$ geometry, with each monomer described by a CAS(8,8)SCF wave function. In particular, the off-diagonal exchange matrix element, $\langle\phi_0^2|V\mathcal{P}_2|\phi_0^1\rangle$, is comparable in magnitude to its electrostatic counterpart, $\langle\phi_0^2|V|\phi_0^1\rangle$, throughout the $\varphi$ rotation. The eigenproblem in Eq.~\eqref{ExchEigenproblem} remains symmetric to within less than 0.2~m$E_h$.

\begin{table}
\centering
\caption{Comparison of dSAPT(CAS) matrix elements appearing in Eq.~\eqref{ExchEigenproblem} for the H$_2$O$\cdots$H$_2$O dimer. The results correspond to $R_{OO}=7.0~a_0$, $\varphi=0^\circ$. Expectation values of the $V$ and $V\mathcal{P}_2$ operators are given in m$E_h$. The dimensionless $\mathcal{P}_2$ matrix elements are scaled by a factor of $10^{3}$.}
\label{MatrixElementsdSRSH2O}
 \begin{tabular}{l r r r r} \hline 
$\hat{O}$ & $\langle\phi_0^1|\hat{O}|\phi_0^1\rangle$ & $\langle\phi_0^2|\hat{O}|\phi_0^2\rangle$ & $\langle\phi_0^1|\hat{O}|\phi_0^2\rangle$ & $\langle\phi_0^2|\hat{O}|\phi_0^1\rangle$ \\ \hline
$V$ & $-4.3$ & $-4.3$ & $-0.9$ & $-0.9$ \\$V\mathcal{P}_2$ & $3.8$ & $3.8$ & $-1.3$ & $-1.3$ \\  $\mathcal{P}_2$ & $-9.0$ & $-9.0$ & $7.5$ & $7.5$ \\ \hline
\end{tabular}
\end{table}

\begin{figure*}
\centering
\includegraphics[width=1\linewidth]{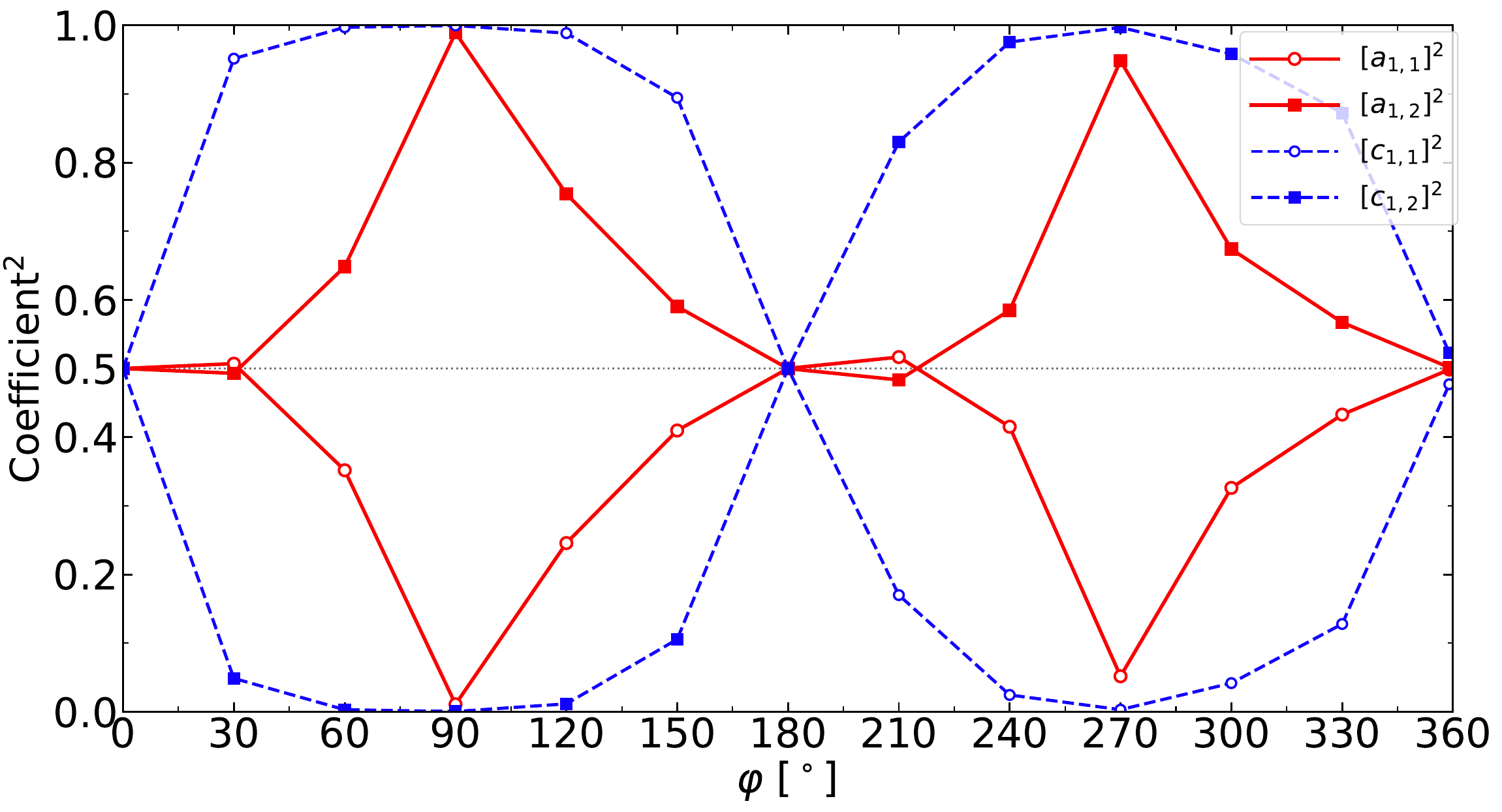}
\caption{Degenerate SAPT(CAS) zeroth-order dimer wave function expansion coefficients for the H$_2$O$\cdots$H$_2$O excimer as a function of the rotation angle $\varphi$ (see text). The $c_{1,i}^2$ and $a_{1,i}^2$ ($i=1,2$) coefficients are obtained from electrostatic [Eq.~\eqref{psi0IJ}] and first-order coupling [Eq.~\eqref{PsiIj0}], respectively.}
\label{fig:h2ocoeff}
\end{figure*}

Figure~\ref{fig:h2ocoeff} shows the angular dependence of the squared zeroth-order wave-function coefficients obtained from the degenerate SAPT(CAS) eigenvalue problems with antisymmetry imposed \textit{after} and \textit{before} diagonalization, i.e., Eqs.~\eqref{matrixElstNEW} and~\eqref{ExchEigenproblem}, respectively. The squared coefficients (occupation numbers) provide a measure of the localization of the excitation on a given monomer. In particular, $[c_{1,1}]^2$ and $[a_{1,1}]^2$ quantify the contribution of the state in which the excitation is localized on the rotating monomer.

At high-symmetry geometries ($\varphi=0^\circ$, $180^\circ$, $360^\circ$), the excitation is equally distributed between the two monomers as a consequence of symmetry and the corresponding state weights are equal to one-half. Specifically, the two monomers are related by reflection in a symmetry plane at $\varphi=0^\circ$ and by inversion at $\varphi=180^\circ$. 
Away from these high-symmetry points, however, electrostatic coupling alone causes the excitation to localize rapidly onto a single monomer (Figure~\ref{fig:h2ocoeff}). Pauli repulsion qualitatively alters this behavior by promoting delocalization throughout the intermediate $C_s$-symmetry region. For instance, at $\varphi=30^\circ$, the polarization approximation predicts an excitation localized almost entirely on the rotating water molecule ($[c_{1,1}]^2=0.95$), whereas dSAPT maintains an even distribution across the complex ($[a_{1,1}]^2=0.51$). Analogous observations follow for a hydrogen-bonded excimer geometry (see Table V in Ref.~\citenum{vanHemert1979abi}).

The contrasting roles of electrostatics and exchange become even more pronounced at $\varphi=90^\circ$ and $\varphi=270^\circ$. Resolving the degeneracy via electrostatics alone [Eq.~\eqref{matrixElstNEW}] predicts an unphysical switch in excitation character: the excitation localizes on the rotating molecule at $\varphi=90^\circ$ ($[c_{1,1}]^2=1$), but transfers to the stationary molecule at $\varphi=270^\circ$ ($[c_{1,1}]^2=0$). Incorporating exchange into the eigenproblem [Eq.~\eqref{ExchEigenproblem}] prevents this state flip, keeping the excitation consistently localized on the stationary monomer at both geometries ($[a_{1,1}]^2=0$).

\begin{figure*}
\centering
\includegraphics[width=1\linewidth]{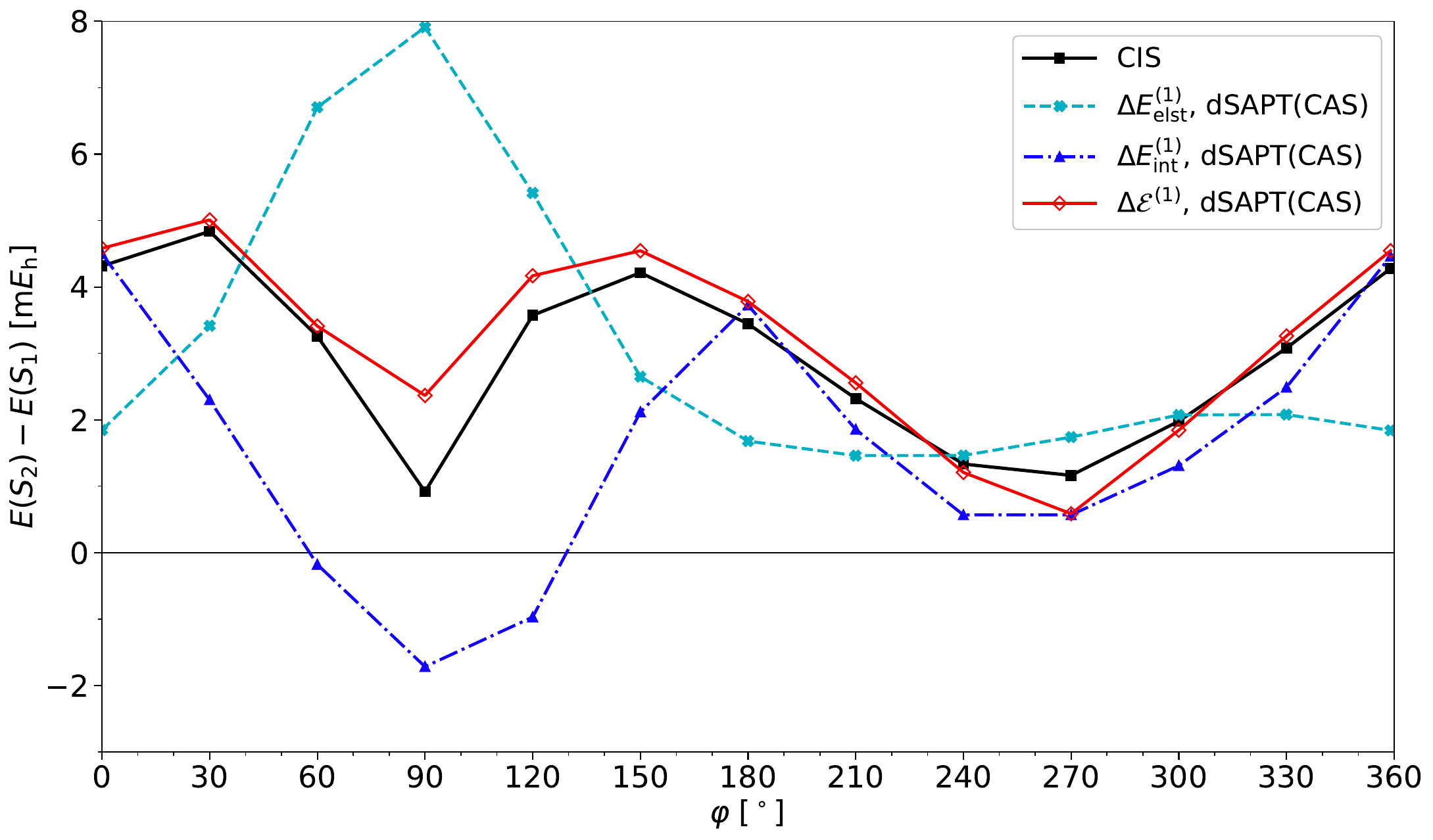}
\caption{Energy difference between the two lowest excited states of the water dimer as a function of the rotation angle $\varphi$ (see the main text for details of the geometry). The EOM-CCSD and CIS energy differences are used as reference values. The quantity $\Delta E^{(1)}_{\mathrm{elst}}$ is obtained as the difference between the two energies resulting from Eq.~\eqref{matrixElstNEW}. The quantity $\Delta E^{(1)}_{\mathrm{int}}$ is evaluated using the same wave-function coefficients and the same state ordering, but with the interaction energies calculated from Eq.~\eqref{E1int}. Finally, $\Delta\mathcal{E}^{\,(1)}$ denotes the difference between the interaction energies obtained from Eq.~\eqref{ExchEigenproblem}. All dSAPT calculation use CAS(8,8)SCF monomers wave functions. }
\label{fig:h2osplit}
\end{figure*}

Although first-order SAPT is generally a poor approximation to the total interaction energy, it may still provide a reasonable estimate of the energy splitting relative to the degenerate non-interacting complex. Figure~\ref{fig:h2osplit} shows the dependence of the energy difference between the two lowest excited states of the water dimer on the rotation angle $\varphi$. The supermolecular CIS results, obtained as the difference between dimer excitation energies, are included as reference data. We verified that the CIS reference closely matches the supermolecular EOM-CCSD benchmark with deviations less than 1~m$E_h$ (see Figure S11 in the Supporting Information). We first consider the splitting obtained from the electrostatic interaction alone, $\Delta E_{\mathrm{elst}}^{(1)}$, defined as the difference between the eigenvalues of Eq.~\eqref{matrixElstNEW}. The resulting curve has a qualitatively incorrect shape, indicating that exchange effects are already significant at the intermolecular separation considered here. Their contribution is particularly pronounced for $\varphi\in(0^\circ,180^\circ)$, where the hydrogen atoms of the rotating monomer point toward the intermolecular region.

When exchange contributions are included \emph{after} diagonalization, that is, when Eq.~\eqref{E1int} is evaluated using the wave-function coefficients and state ordering obtained from the electrostatic eigenvalue problem [Eq.~\eqref{matrixElstNEW}], the resulting splitting, $\Delta E_{\mathrm{int}}^{(1)}$, improves significantly and becomes qualitatively correct (Figure~\ref{fig:h2osplit}). However, in the region where exchange effects are strongest, $\varphi\in(0^\circ,180^\circ)$, the splitting is substantially underestimated and even changes sign in the vicinity of $\varphi=90^\circ$. The origin of this failure can be traced to the change in the character of the lower-energy state induced by antisymmetrization. As discussed above, depending on whether antisymmetrization is imposed \emph{after} or \emph{before} diagonalization, the excitation in the lower-energy state is localized on different monomers. In the relevant $\varphi=90^\circ$ geometry, one obtains $[c_{1,1}]^2=1$ for the electrostatic eigenproblem and $[a_{1,1}]^2=0$ for the exchange-including eigenproblem, respectively (see Figure~\ref{fig:h2ocoeff}).

Finally, incorporating antisymmetrization \emph{before} diagonalization yields the dSAPT(CAS) splitting, $\Delta\mathcal{E}^{(1)}$, which closely follows the supermolecular CIS results (see Figure~\ref{fig:h2osplit}). The largest difference occurs at $\varphi=90^\circ$ and does not exceed $1.5$~m$E_h$. dSAPT calculations employing CIS monomer wave functions give nearly identical results, as shown in Figure~S11 of the Supporting Information.

\begin{figure*}
\centering
\includegraphics[width=1\linewidth]{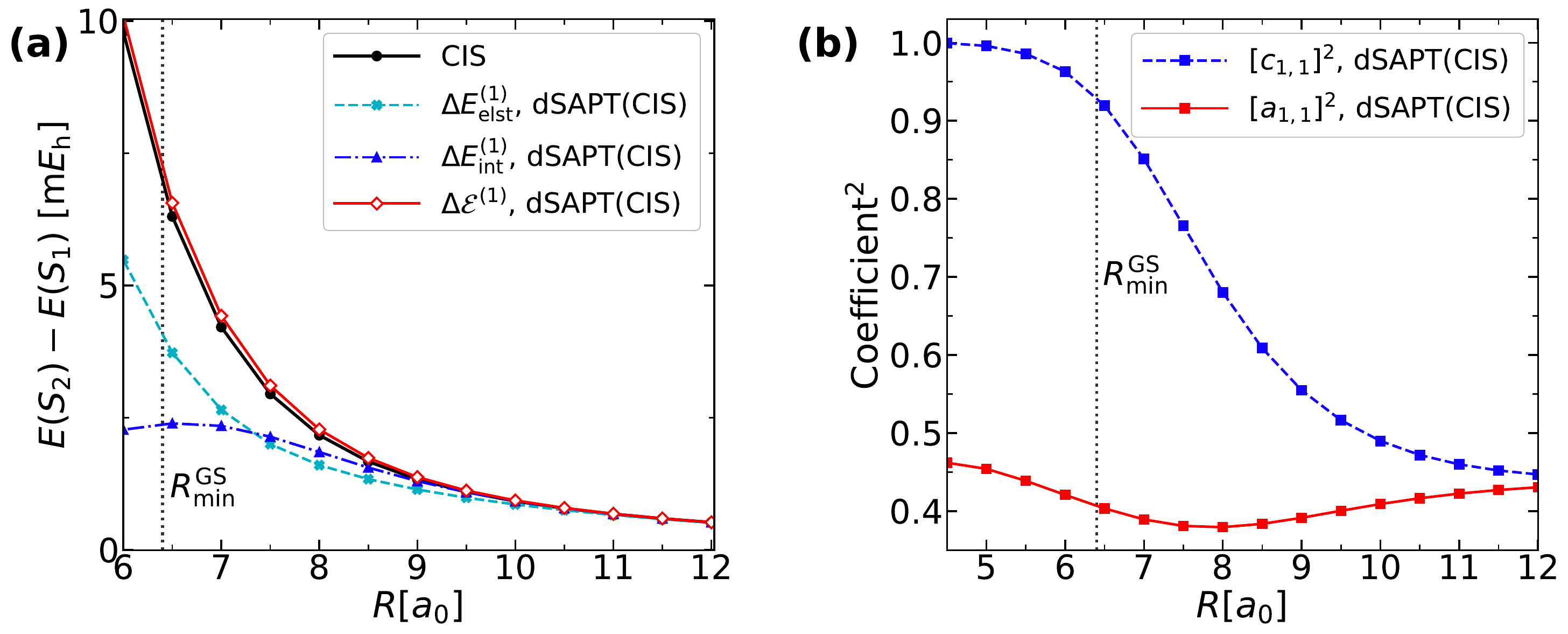}
\caption{Distance $R$ dependence of (a) Energy difference between two lowest excited states and (b) squared coefficient for water dimer for $\varphi=150^\circ$ (see the main text for details of the geometry).  The quantity $\Delta E^{(1)}_{\mathrm{elst}}$ is obtained as the difference between the two energies resulting from Eq.~\eqref{matrixElstNEW}. The quantity $\Delta E^{(1)}_{\mathrm{int}}$ is evaluated using the same wave-function coefficients and the same state ordering, but with the interaction energies calculated from Eq.~\eqref{E1int}. Finally, $\Delta\mathcal{E}^{\,(1)}$ denotes the difference between the interaction energies obtained from Eq.~\eqref{ExchEigenproblem}. The $c_{1,i}^2$ and $a_{1,i}^2$ ($i=1,2$) coefficients are obtained from electrostatic [Eq.~\eqref{psi0IJ}] and first-order coupling [Eq.~\eqref{PsiIj0}], respectively. The $R_{\rm min}$ distance represents the minimum of the ground state interaction energy with respect to $R$ calculated at the SAPT0 level. All dSAPT calculation use CIS monomer wave functions.}
 \label{fig:h2oRdep}
\end{figure*}

Let us examine how the energy splitting varies with the intermolecular distance $R$. In Figure~\ref{fig:h2oRdep}, we illustrate this dependence for a low-symmetry geometry corresponding to $\varphi=150^\circ$. 
As shown in panel (a), at large intermolecular separations ($R>10\,a_0$), exchange effects are negligible, and all SAPT variants agree well with the reference supermolecular CIS curve. As $R$ decreases, exchange contributions grow, and the splitting obtained from the electrostatic interaction alone, $\Delta E_{\mathrm{elst}}^{(1)}$, diverges from the other results. Below $R=8.5\,a_0$, applying antisymmetrization \emph{after} diagonalization ($\Delta E_{\rm int}^{(1)}$) fails because it neglects the exchange-induced relaxation of the zeroth-order wave-function coefficients. In contrast, the fully antisymmetrized dSAPT(CAS) splitting $\Delta\mathcal{E}^{(1)}$ reproduces the reference supermolecular CIS curve across the entire range of distances.

Panel (b) of Figure~\ref{fig:h2oRdep} illustrates the $R$-dependence of the squared coefficients. Here, we compare coefficients obtained with antisymmetrization imposed \emph{after} and \emph{before} diagonalization. While both approaches converge asymptotically to approximately 0.44 at long range, they diverge sharply as the monomers approach each other. As observed for the angular dependence (Figure~\ref{fig:h2ocoeff}), electrostatics alone artificially localizes the excitation onto a single monomer at short distances, whereas coupling via both electrostatics and electron exchange keeps the excitation delocalized, leading to an almost equal distribution of the excitation at short intermolecular distances. This confirms that exchange effects must be included directly in the diagonalization procedure in order to obtain quantitatively accurate energy splittings.

\subsection{The benzene (T-shaped) excimer}
The benzene dimer is a canonical model of $\pi$--$\pi$ interactions and one ofthe prototypical systems for studying excimer formation \cite{azumi1965ene,rocharinza2006ath,huenerbein2008tim,ge2018ene}.
Here, we restrict the analysis to geometries that can be  directly probed experimentally. Jet-cooled spectroscopy of the benzene dimer selectively probes configurations with non-negligible transition dipole moment, i.e.\ the distorted, tipped T-shape configuration. More symmetric structures, including parallel-displaced and stacked geometries, are essentially optically dark. At the near-T-shaped geometry, this splitting is dominated by the inequivalence of the two molecular sites rather than by a pure excitonic coupling. This makes the benzene dimer a demanding test case for a degenerate perturbative treatment.

The reference T-shaped (C$_{\rm 2v}$) geometry was taken from Ref.~\cite{balmer2015the} corresponding to a ground-state saddle point (S3 in the nomenclature of Ref.~\citenum{podeszwa2006pot}). In dSAPT(CAS) calculations, each monomer was described using a SA(2)-CAS(6,6)SCF wave function. As reference, we adopted vertical excitations from spin-component scaled (SCS) CC2 method which was shown to closely match the experimentally observed excited-state splitting \cite{balmer2015the}. The SCS-CC2 calculations were performed in the MRCC program \cite{kallay2020,mester2025} with scaling factors as proposed by Grimme \cite{grimme2003imp}. All calculations employed the aug-cc-pVTZ basis set.

\begin{figure*}
\centering
\includegraphics[width=1\linewidth]{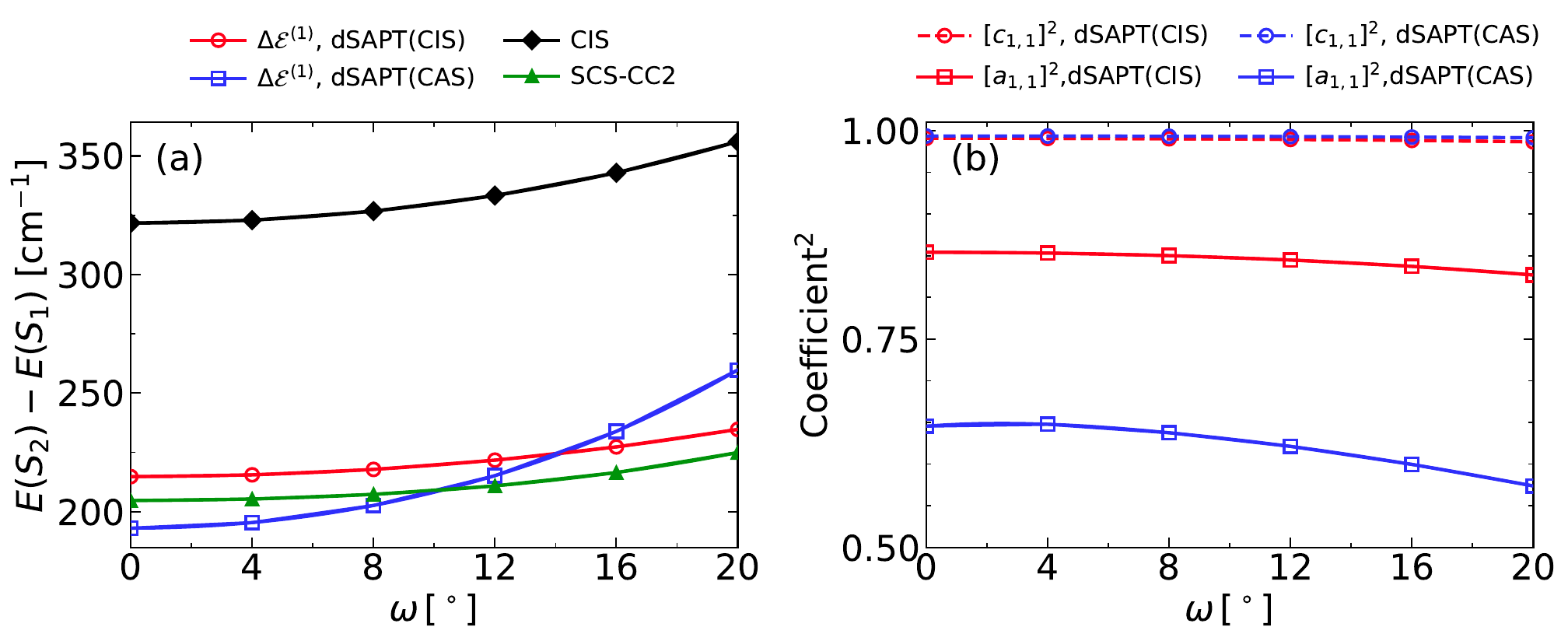}
\caption{Energy difference between the two lowest excited states of the benzene dimer as a function of the tipping angle $\omega$ (see the main text for details of the geometry). The $c_{1,i}^2$ and $a_{1,i}^2$ ($i=1,2$) coefficients are obtained from electrostatic [Eq.~\eqref{psi0IJ}] and first-order coupling [Eq.~\eqref{PsiIj0}], respectively. The SCS-CC2 and CIS energy differences are used as reference values.}
\label{fig:benzene}
\end{figure*}

Figure~\ref{fig:benzene} shows the energy splitting between the $S_1$ and $S_2$ states of the benzene dimer as a function of the tipping angle $\omega$. The scan connects the S3 saddle point structure at $\omega=0^\circ$ with the geometry close to the ground-state global miniumum at $\omega=20^\circ$, see Figure~\ref{geoms}. Since the geometries along the scan were not reoptimized, the $\omega=20^\circ$ does not coincide exactly with the true global minimum.

As evident from Figure~\ref{fig:benzene}(a), the splitting depends only weakly on the tipping angle. At the SCS-CC2 level, it varies between 205 and 225~cm$^{-1}$ which matches the experimentally observed \cite{balmer2015the} value of ca.\ 250~cm$^{-1}$. The first-order dSAPT(CIS) curve is nearly parallel to the SCS-CC2 reference and systematically overestimates the splitting by only 9-11~cm$^{-1}$. This behavior contrasts with CIS results, which overestimate the energy gap by roughly 120~cm$^{-1}$. The dSAPT(CAS) splittings span a wider range, approximately 190-260~cm$^{-1}$, and thus show a stronger angular dependence than the SCS-CC2 benchmark. Nevertheless, the agreement remains satisfactory: dSAPT(CAS) slightly underestimates the splitting in the T-shaped geometry and overestimates it near the tilted ground-state minimum. In contrast, nondegenerate first-order SAPT(CAS) separates the states by merely 30~cm$^{-1}$.

The squared coefficients in Figure~\ref{fig:benzene}(b) provide further insight into the relative roles of electrostatics and Pauli exchange in configuration mixing.  When the degeneracy is resolved through electrostatics alone, the excitation is almost completely localized on the ``stem'' benzene: the $[c_{1,1}]^2$ coefficients remain close to unity throughout the $\omega$ scan. Consistently, the $\Delta E^{(1)}_{\rm int}$ splitting obtained with antisymmetrization imposed \emph{after} diagonalization is essentially identical to the first-order nondegenerate SAPT splitting. As in the water excimer, partial delocalization appears only when Pauli exchange is included directly in the first-order eigenproblem. In the dSAPT(CIS) picture, this effect is relatively weak and the excitation remains largely localized on the same monomer ($[a_{1,1}]^2=0.82$--$0.85$). In contrast, dSAPT(CAS) predicts a more pronounced delocalization ($[a_{1,1}]^2=0.58$--$0.65$). 

\section{Conclusions}

In this work, we formulated first-order degenerate SAPT within SRS theory. The resulting framework enables the treatment of orbital and resonance degeneracies arising between weakly coupled subsystems. We compared two distinct ways of resolving the degeneracy at first-order. In the first approach, antisymmetry is imposed \textit{after} diagonalization of the effective electrostatic interaction. In the second approach, the antisymmetrizer in incorporated already in the zeroth-order wave function, \textit{before} diagonalization of the first-order eigenproblem. Thus, in the former case the coupling between degenerate states is determined solely by electrostatics, whereas in the latter case exchange effects also contribute to the zeroth-order wave function. The correspoding first-order exchange energies are formally related only in the small-overlap region.

To assess the importance of Pauli repulsion for configuration mixing, we studied two prototypical open-shell van der Waals complexes, F(${}^2$P)$\cdots$H$_2$ and NO($^2\Pi$)$\cdots$H$_2$, with monomers described at the CASSCF level. An important advantage of degenerate SAPT is that it avoids the need for diabatization of supermolecular adiabatic states, since the interaction matrix is constructed directly in the diabatic basis spanning the degenerate zeroth-order subspace. As expected, electrostatics alone yields reliable mixing angles only in the long-range regime, approximately for $R>10~a_0$. In both systems, exchange effects become important already at intermediate intermolecular separations. For F$\cdots$H$_2$, inclusion of Pauli repulsion gives a qualitatively correct description in the ver der Waals minimum, with the remaining discrepancies relative to supermolecular CASSCF results attributed primarily to induction effects. For NO$\cdots$H$_2$, the angular dependence of the mixing angle is more intricate. First-order dSAPT(CAS) remains in excellent agreement with the coupled-cluster reference \cite{deJongh2017ima} even in the repulsive wall region of T-shaped geometries. For skewed H-shaped configurations, it still substantailly improves upon the purely electrostatic picture, although second-order effects begin to play an important role. 

For both two-state systems with orbital degeneracy, we also evaluated second-order energy contributions using a diabatization procedure based on single-reference SAPT  with unrestricted Hartree-Fock or Kohn-Sham monomer wave functions. This analysis showed that the second-order dispersion energy can significantly modify state mixing, especially in skewed H-shaped geometries. 
This finding motivates the future extension of degenerate SAPT beyond first order.

Until now, SAPT treatment of excited-state complexes has been limited to cases in which the excitation is localized on a single monomer \cite{jangrouei2022dis,hapka2023eff,krzeminska2024ani}. Degenerate SAPT extends this scope to excitations delocalized over the entire system. Using a simple water excimer model probing both the angular and distance dependence, we demonstrated that Pauli exchange can strongly affect the degree of exciton delocalization. At the high-symmetry geometries, the mixing is fixed by symmetry, which enforces the equal-weight excitonic combinations
$\frac{1}{\sqrt{2}}\left( \ket{AB^*} \pm \ket{A^*B} \right)$. In the purely electrostatic picture, the excitation localizes as soon as the complex deviates from these high-symmetry solutions and may switch between the monomers depending on their relative orientation. In contrast, including exchange effects promotes delocalization and stabilizes the exciton as one of the molecules is rotated. We verified this behavior by examining the splitting between asymptotically degenerate excited states as function of angle and distance. The polarization approximation, which accounts for electrostatics alone, is qualitatively incorrect in the degenerate case. In contrast, dSAPT(CAS) and dSAPT(CIS) closely recover the supermolecular CIS and EOM-CCSD reference results. These findings are corroborated by the excited benzene dimer. In the tipped T-shaped arrangement, which has been characterized by electronic spectroscopy \cite{balmer2015the}, Pauli repulsion again enables exciton delocalization and brings the $S_1$-$S_2$ splitting to the correct value.

To summarize, among the two possible first-order dSAPT formulations, only the
approach in which antisymmetrization is imposed \textit{before} diagonalization includes Pauli-repulsion effects directly in the zeroth-order wave functions. The resulting first-order dSAPT diabatization protocol may therefore prove useful for modeling interaction potentials in the long- and intermediate-range regimes of collision processes involving open-shell van der Waals dimers. One should bear in mind, however, that this formulation requires solving a non-Hermitian eigenproblem. Consequently, deviations from the expected symmetry relations should always be monitored in practical calculations. By contrast, antisymmetrization \textit{after} diagonalization leads to a Hermitian eigenproblem and preserves a rigorous separation between electrostatic and exchange energies, but does not incorporate exchange effects into the zeroth-order state mixing. The extension of dSAPT to second order in the intermolecular interaction operator is currently under development in our groups.

\section{Appendix I}
Here, we present an alternative approach to resolving the spatial degeneracy using only nondegenerate SAPT. First, introduce parametrization of a degenerate monomer state via the angle $\varphi$,
\begin{equation}
    \ket{A(\varphi)} 
    = \cos(\varphi)\, \ket{A_{0_1}}
    + \sin(\varphi)\, \ket{A_{0_2}} \, .
\end{equation}
The key assumption is that this state can be expressed as a single determinant constructed from rotated orbitals. Using the F(${}^2$P)$\cdots$H$_2$ example, we get 
\begin{equation*}
    |A(\varphi)\rangle 
    = |1s^2 2s^2 2p_y^2\, [2p_t(\varphi)]^2 [2p_t^{\perp}(\varphi)]^1\rangle.
\end{equation*}
where the $2p_t$ and $2p_t^{\perp}$ orbitals are defined as
\begin{equation}
\begin{split}
2p_t(\varphi)         &= \cos(\varphi)\,2p_x + \sin(\varphi)\,2p_z,  \\
2p_t^{\perp}(\varphi) &= -\sin(\varphi)\,2p_x + \cos(\varphi)\,2p_z.
\end{split}
\end{equation}
Next, construct a product state $|\phi(\varphi)\rangle=\ket{A(\varphi) B}$, which serves as a zeroth-order wave function for nondegenerate SAPT(UHF) calculation. Finally, the mixing angle can be determined from
\begin{align}\label{SAPT(UHF)elst}
\gamma_{\text{elst}} &= \text{arg\,min}_\varphi [E^{(1)}_{\text{elst}}(\varphi)] \\ \label{SAPT(UHF)exch}
\gamma_{\text{1st.ord}} &= \text{arg\,min}_\varphi [E^{(1)}_{\text{elst}}(\varphi)+E^{(1)}_{\text{exch}}(\propto S^2)(\varphi)]
\end{align}
Here, we use the fact that for a $2 \times 2$ symmetric matrix $\mathbb{A}$, finding its eigenvector is equivalent to finding a unit vector $|v\rangle$ (i.e., $\langle v|v\rangle = 1$) that  minimizes the quantity $\langle v|\mathbb{A}|v\rangle$. The same procedure can also be generalized to resolve degeneracy in higher orders of SAPT. In this way, the mixing angle is determined without explicitly calculating the off-diagonal matrix element of the (symmetry-adapted) interaction operator within the degenerate subspace.


\begin{acknowledgements}
We thank E.\ Giner for his help with the implementation of transition density matrices in Quantum Package. The National Science Center of Poland supported this work under grant no.\ 2021/43/D/ST4/02762 (MH and DC) and no.\ 2019/34/E/ST4/00407 (PSZ, Sonata Bis program). For the purpose of Open Access, the author has applied a CC-BY public copyright license to any Author Accepted Manuscript (AAM) version arising from this submission.
\end{acknowledgements}

\bibliography{articles}   

\end{document}